\documentclass[fleqn,usenatbib]{mnras}

\usepackage{newtxtext,newtxmath}

\usepackage[T1]{fontenc}

\DeclareRobustCommand{\VAN}[3]{#2}
\let\VANthebibliography\thebibliography
\def\thebibliography{\DeclareRobustCommand{\VAN}[3]{##3}\VANthebibliography}

\usepackage{graphicx}	
\usepackage{amsmath}	

\usepackage{color}

\usepackage{ulem} 
\usepackage{url}
\usepackage{threeparttable} 
\usepackage[figuresright]{rotating} 
\usepackage{booktabs}
\usepackage{array}
\usepackage{makecell}
\usepackage{multirow}
\usepackage{pdflscape}
\usepackage{setspace}

\title[A variable corona in MAXI J1820+070]{A variable corona during the transition from type-C to type-B quasi-periodic oscillations in the black hole X-ray binary MAXI J1820+070}

\author[Ruican Ma et al.]{
Ruican Ma$^{1,2,3}$\thanks{E-mail: maruican@astro.rug.nl},
Mariano M\'endez$^{1}$\thanks{E-mail: mariano@astro.rug.nl},
Federico Garc\'ia$^{4}$\thanks{E-mail: fgarcia@iar.unlp.edu.ar},
Na Sai,$^{5,6}$
Liang Zhang,$^{2}$
Yuexin Zhang$^{1}$
\\
$^{1}$Kapteyn Astronomical Institute, University of Groningen, P.O. BOX 800, 9700 AV Groningen, The Netherlands\\
$^{2}$Key Laboratory of Particle Astrophysics, Institute of High Energy Physics, Chinese Academy of Sciences, Beijing 100049, China\\
$^{3}$University of Chinese Academy of Sciences, Chinese Academy of Sciences, Beijing 100049, China\\
$^{4}$Instituto Argentino de Radioastronom\'ia (CCT La Plata, CONICET; CICPBA; UNLP), C.C.5, (1894) Villa Elisa, Buenos Aires, Argentina\\
$^{5}$Department of Astronomy, School of Physics and Technology, Wuhan University, Wuhan 430072, China\\
$^{6}$WHU-NAOC Joint Center for Astronomy, Wuhan University, Wuhan 430072, China
}

\date{Accepted 2023 July 24. Received 2023 July 24; in original form 2022 December 14}

\pubyear{2022}

\begin{document}
\label{firstpage}
\pagerange{\pageref{firstpage}--\pageref{lastpage}}
\maketitle

\begin{abstract}
We analyze a \textit{Neutron Star Interior Composition Explorer} (NICER) observation of the black hole X-ray binary MAXI J1820+070 during a transition from type-C to type-B quasi-periodic oscillations (QPOs). We find that below $\sim$2\,keV, for the type-B QPOs the rms amplitude is lower and the magnitude of the phase lags is larger than for the type-C QPOs. Above that energy, the rms and phase-lag spectra of the type-B and type-C QPOs are consistent with being the same. We perform a joint fit of the time-averaged spectra of the source, and the rms and phase-lag spectra of the QPOs with the time-dependent Comptonization model {\tt vkompth} to study the geometry of the corona during the transition. We find that the data can be well-fitted with a model consisting of a small and a large corona that are physically connected. The sizes of the small and large coronae increase gradually during the type-C QPO phase whereas they decrease abruptly at the transition to type-B QPO. At the same time, the inner radius of the disc moves inward at the QPO transition. Combined with simultaneous radio observations showing that discrete jet ejections happen around the time of the QPO transition, we propose that a corona that expands horizontally during the type-C QPO phase, from $\sim10^{4}$\,km ($\sim800\,R_{\rm g}$) to $10^{5}$\,km ($\sim8000\,R_{\rm g}$) overlying the accretion disc, transforms into a vertical jet-like corona extending over $\sim10^{4}$\,km ($\sim800\,R_{\rm g}$) during the type-B QPO phase.

\end{abstract} 




\section{Introduction}

Most black-hole binaries (BHBs) are transient sources that, after months or even decades in a quiescent state, show an X-ray outburst that typically lasts weeks to months, driven by mass accretion from the secondary star onto the black-hole primary \citep{Tanaka1996}. For a typical black hole transient (BHT), the emission in the X-ray band is mainly produced by two components in the accretion flow: At low energies, the spectrum is dominated by a geometrically thin, optically thick accretion disc \citep{Shakura1973} that produces a blackbody-like thermal spectrum with a characteristic temperature around 1.0\,keV \citep[for a review, see][]{McClintock2006}. At high energy the spectrum is dominated by emission from a plasma of highly-energetic electrons, the so-called X-ray corona, in which the disc photons are inverse-Compton scattered producing a power-law-like component extending up to 100\,keV \citep[e.g.,][]{Sunyaev1980}. A reflection component from fluorescence, photoabsorption, and Compton scattering due to the irradiation of the accretion disc by the hot corona  \citep{Basko1974, George1991} is also observed in some sources \citep[e.g., ][]{Xu2018, Chakraborty2021, LiuHH2022}. In the time-averaged spectrum, the main features of this component are a broad Fe K$_{\rm \alpha}$ line at 6.4--7\,keV, a Fe K edge at 7--10\,keV and a broad Compton hump at 20--30\,keV \citep[see ][for a review]{Gilfanov2010}.

A ``canonical'' outburst of a BHT is divided into different X-ray spectral states \citep[e.g.,][]{Mendez1997} as the source follows a counter-clockwise ``q'' shape in the hardness intensity diagram \citep[HID; e.g.,][]{Homan2001, Homan2005, Belloni2016}. At the beginning of the outburst, the source is in the low hard state (LHS), the time-averaged spectrum is dominated by the non-thermal component with a power-law index $\Gamma\sim$1.5. Both strong low-frequency band-limited noise components and low-frequency quasi-periodic oscillations \citep[QPOs;][]{vdKlis1985, vdKlis2006} usually appear in this state (see below). In this scenario, in the LHS the accretion disc is truncated, with a truncation radius of tens to hundreds of $R_{\rm g}$ \citep{McClintock2001, Done2007, Done2010}. However, the detection of broad iron lines \citep[e.g.,][]{Miller2006} and residual thermal disc emission \citep[e.g.,][]{Rykoff2007} in the LHS suggest that the disc remains at or near the innermost stable circular orbit \citep[ISCO;][]{Esin1997}. The radio emission coming from a compact, persistent jet also exists in this state \citep[e.g.,][]{Fender2001}. As the luminosity increases, the inner radius of the accretion disc moves inward \citep{Gierlinski2008}, the disc temperature increases and the emission of the disc component becomes comparable to that of the non-thermal component \citep{Cuneo2020}, indicating that the source enters the intermediate state \citep[IMS; see][and references therein]{Homan2005,McClintock2006}. The IMS can be further divided into the hard intermediate (HIMS) and soft intermediate states (SIMS), with the spectrum of the source being softer in the SIMS than in the HIMS \citep[][]{Homan2005}. The strong broad-band noise and QPOs in the LHS and HIMS are replaced by a relatively weaker broad-band noise and QPO components in the SIMS \citep[][]{Homan2005}. The compact jet is quenched and then a highly relativistic and discrete jet is launched in the IMS \citep{Fender2004}. When the source enters the high-soft state (HSS), the thermal accretion-disc component dominates the time-averaged spectrum, and the power-law index of the non-thermal component increases to $\Gamma>$2.1 \citep[][]{McClintock2006}. A weak QPO together with weak broadband noise sometimes appears in this state \citep{Motta2016}. In the HSS the jet turns off and the inner radius of the disc reaches the ISCO. As the flux of the source decreases, the source returns to the intermediate state and the low-hard state where the compact jet reappears \citep[][]{McClintock2006}, and finally, the source goes back into quiescence.

\subsection{Low-frequency quasi-periodic oscillations}
The fast X-ray variability in BHTs has been widely studied as an important characteristic of the source during outbursts. Quasi-periodic oscillations (QPOs) are one of the most prominent features observed in these sources during the outburst \citep[e.g.,][]{vdKlis1985, vdKlis2006}, which appear as one or more narrow peaks in the power density spectrum (PDS) of the source \citep[\citealt{vdKlis1989}; see][for a review]{Ingram2019}. When the centroid frequency is between mHz and $\sim$30\,Hz, the QPOs are classified as low-frequency QPOs \citep[LFQPOs;][]{Motta2016}. According to the shape and strength of the noise component in the PDS and the root-mean-square (rms) amplitude and phase lags of the QPOs, LFQPOs are classified as type-A, -B or -C QPOs \citep{Wijnands1999, Remillard2002, Casella2005}. Type-C QPOs are the most common and strongest QPOs in BHTs and are generally observed in the LHS and HIMS. The rms amplitude of type-C QPOs can reach up to 20\%, with a high-quality factor, $Q\geq$10 ($Q=\nu/{\rm FWHM}$, where $\nu$ and FWHM are the centroid frequency and the full width at half maximum of the QPO, respectively). Type-C QPOs usually show subharmonics, second and sometimes third harmonics, and are accompanied by a strong broadband noise component in the PDS. 

The models proposed to explain the type-C QPOs are roughly divided into two categories. One class of models explains the type-C QPO as either some sort of instability in the accretion flow, e.g, an accretion-ejection instability in a magnetized disc \citep[][]{Tagger1999}, oscillations in a transition layer in the accretion flow \citep[][]{Titarchuk2004}, or corona oscillations caused by magneto-acoustic waves \citep{Cabanac2010}. The other class of models explains the QPO as a geometric effect under general relativity. The most popular of this class of models is the Lense-Thirring (LT) model proposed by \citet{Stella1998}. In particular, \citet[][see also \citealt{You2018}]{Ingram2009} developed the LT model considering a precessing accretion flow inside a truncated accretion disc to explain the evolution of the spectral parameters of the source as a function of the type-C QPO cycle \citep{Ingram2016}.  

Type-B QPOs are weaker than type-C QPOs and appear in the SIMS (in fact, the appearance of type-B QPOs is one of the characteristics that defines the start of the SIMS). These QPOs have rms amplitudes of $\sim$4\%, $Q\geq$6, and are accompanied by a weak low-frequency noise component. One of the explanations for type-B QPOs is that they originate from the jet \citep[e.g., ][]{Soleri2008, Stevens2016, deRuiter2019, LiuHX2022}. Type-A QPOs are the weakest LFQPOs, they appear in the HSS and are rarely detected \citep[][]{Motta2016}. 

Transitions between different types of QPOs have been reported in several BHTs. QPO transitions in GX 339--4 in four outbursts were systematically studied by \citet{Motta2011}, who suggested that type-C and type-A QPOs have different origins than type-B QPOs. The transition from type-B to type-A QPOs in XTE J1550--564 is associated with an increase in the count rate and discrete ejections in the radio jet \citep{Homan2001, Sriram2016}. However, the same type-B to type-A QPO transition in XTE J1817--330 is connected to a flux decrease \citep{Sriram2012}. XTE J1859+262 shows a variety of changes in the parameters of the accretion disc during the type-B to type-A QPOs transition \citep{Sriram2013}, whereas the transition from type-C to type-A and type-B QPOs in this source appears to be related to relativistic ejection episodes \citep{Casella2004}. \citet{LiuHX2022} reported that the transition between type-B and type-C QPOs in MAXI J1348--630 is associated with an increase of the Comptonized flux, the hardening of the energy spectrum, and a significant change of the inner radius of the accretion disc. \citet{Soleri2008} reported the transition between type-C and type-B QPOs in GRS 1915+105 and proposed that the type-B QPOs are related to discrete jet ejections in the SIMS. Recently \citet{Homan2020} reported the transition from type-C to type-B QPOs in MAXI J1820+070 and provided strong evidence that the appearance of type-B QPOs is related to discrete jet ejections. 

Phase lags give extra information about the X-ray variability in BHTs. The phase lags of the variability are measured using the Fourier cross spectrum computed from light curves in two different energy bands \citep{Miyamoto1989,Cui1997,Nowak1999}. There are a variety of mechanisms proposed to explain the phase lags. For example, hard lags, in which high-energy photons are delayed with respect to the low-energy ones, could originate from the Comptonization of soft disc photons in the corona \citep{Payne1980,Kazanas1997}. Alternatively, accretion rate fluctuations in the accretion disc that propagate inward could also explain the observed hard phase lags \citep{Lyubarskii1997, Kotov2001}. Soft phase lags, on the other hand, could be produced when hard photons from the corona irradiate the accretion disc and are reprocessed and re-emitted at a later time than the corona photons that go directly to the observer \citep[e.g., ][]{Uttley2014,Ingram2019}. 

\subsection{Geometry of the corona from the energy-dependent variability}
The geometry of the accretion flow near the black hole in these systems remains a topic of debate, and several accretion disc-corona coupling models have been proposed \citep[e.g., ][]{Reig2003, Ingram2009, Kylafis2018, Marcel2018,  Zdziarski2021, Kawamura2022, Mastichiadis2022}. 

Recently, \citet{Karpouzas2020} and \citet{Bellavita2022} developed a time-dependent Comptonization model that addresses this coupling. These models are based on the ideas originally proposed by \citet{Lee1998}, \citet{Lee2001} and \citet{Kumar2014}. In this model the QPO arises from the coupled oscillation between the corona and the disc that leads to the oscillation of the physical parameters that describe the spectrum of the system: the temperature of the corona, $kT_{\rm e}$, the temperature of the seed-photons from the (disc) blackbody, $kT_{\rm s}$, and the external heating rate that keeps the corona in thermal equilibrium, $\Dot{H}_{\rm ext}$. In addition, these models give the corona size, $L$, and the feedback fraction, $0 \leq \eta \leq 1$, which is defined as the fraction of the flux of the disc due to feedback from the corona. The feedback fraction, $\eta$, is related to the fraction of the flux of corona photons that return to the disc, $\eta_{\rm int}$\footnote{The intrinsic feedback fraction is defined as the ratio of the flux of the corona that illuminates the disc to the total corona flux.} \citep[intrinsic feedback fraction; see][]{Karpouzas2020}. Based on the above parameters, we can obtain information on the geometry of the corona. This model is generically called {\tt vkompth}\footnote{The {\sc vkompth} model is publicly available at \url{https://github.com/candebellavita/vkompth}.}, and has been applied to the QPOs in a number of X-ray binaries to place constraints on the corona geometry of these systems. For example, the model was applied successfully to the kilohertz (kHz) QPOs in the neutron-star system 4U 1636$-$53 \citep{Karpouzas2020}, the type-B QPO in the black-hole system MAXI J1348$-$630 \citep{Garcia2021, Bellavita2022}, GX 339--4 \citep{Peirano2022b}, MAXI J1535--571 \citep{Zhang2023}, and the type-C QPOs in GRS 1915+105 \citep{Karpouzas2021, Garcia2022, Mendez2022} and MAXI J1535$-$571 \citep{Zhang2022,Rawat2023}. Here we apply this time-dependent Comptonization model to the type-C and type-B QPOs in the black-hole system MAXI J1820+070 in the observation presented in \citet{Homan2020}.

\subsection{MAXI J1820+070}
MAXI J1820+070 is a BHT \citep{Tucker2018} that exhibited a rich variety of spectral and timing properties during its 2018 outburst. MAXI J1820+070 was discovered with the \textit{All-Sky Automated Survey for SuperNovae} \citep[ASAS-SN;][]{Shappee2014,Kochanek2017} as the optical transient ASASSN-18ey on 2018 March 6 \citep[MJD 58183;][]{Denisenko2018,Tucker2018}. Around five days later, this source was observed in the X-ray band with the \textit{Monitor of All-sky X-ray Image} Gas Slit Camera \citep[\textit{MAXI}/GSC;][]{Matsuoka2009} on 2018 March 11 \citep[MJD 58188;][]{Kawamuro2018}. The position of the source is R.A. (J2000) $=18^{\rm h}$20$^{\rm m}$21$^{\rm s}$.94, Dec. (J2000) $=+07^{\circ}$11$^{\prime}$07$\farcs$19 \citep{Gandhi2019}. MAXI J1820+070 is a dynamically confirmed BH with a mass of $8.48^{+0.79}_{-0.72}$\,$M_{\rm \odot}$ and an inclination of $63^{\circ}\pm{0.3^{\circ}}$ \citep{Torres2020} at a distance of $2.96\pm{0.33}$\,kpc \citep{Atri2020}. The source was observed at several wavelength bands, such as radio \citep[e.g., ][]{Atri2020, Homan2020, Wood2021}, NIR \citep[e.g., ][]{Ozbey2022, Yoshitake2022}, optical \citep[e.g., ][]{Paice2021, Thomas2022}, UV \citep[e.g., ][]{Kajava2019, Ozbey2022}, and X-ray \citep[e.g., ][]{Buisson2019, Kara2019, Bright2020, Wang2020, Ma2021, You2021}. After being in the hard state for about 110 days, MAXI J1820+070 transitioned to the intermediate state from MJD 58303.5 to MJD 58310.7 \citep[][]{Shidatsu2019}. During this period there was a switch from type-C to type-B QPOs, accompanied by a flare in the 7--12\,keV band \citep{Homan2020} and a radio flare associated to a superluminal ejection \citep{Bright2020}. The reports of \citet{Homan2020} and \citet{Wood2021} suggest that the ejection is most likely linked to the QPO transition.

In this paper, we use the time-dependent Comptonization model {\tt vkompth} to explore the evolution of the coronal geometry of MAXI J1820+070 during the transition from type-C to type-B QPO. The paper is structured as follows: We describe the observations and data reduction in Section~\ref{sec:data_reduced}, where we also describe the model and parameters used to fit the time-averaged spectra of the source and the rms and phase-lag spectra of the QPOs. We present the results of the joint fitting of all these spectra for the intervals with type-C and type-B QPOs as a function of time and frequency in Section~\ref{sec:results}. In Section~\ref{sec:discussion}, we discuss the corona geometry during the QPO transition. Finally, we summarise our conclusions in Section~\ref{sec:conclusion}.\\

\section{Observation and data analysis}
\label{sec:data_reduced}

The \textit{Neutron Star Interior Composition Explorer} \citep[\textit{NICER;}][]{Gendreau2016} is an X-ray telescope on board the \textit{International Space Station} (\textit{ISS}). The primary scientific instrument of \textit{NICER} is the X-ray Timing Instrument (XTI), consisting of 56 X-ray concentrator optics (XRC), covering the 0.2$-$12\,keV energy range with $\sim$100\,ns time resolution. \textit{NICER} observed the outburst of MAXI J1820+070 from 2018 March 6 (MJD 58183) to 2018 November 21 (MJD 58443). 

We process the observations with the \textit{NICER} data analysis software (NICERDAS) version 2021-09-01\_V008c, using the CALDB version xti20210707. We reprocess the data using \textit{nicerl2} with the standard filtering criteria \citep[see, e.g,][]{Wang2021} and estimate the background using \textit{nibackgen3C50} and the 3C50 model \citep[][]{Remillard2022}. 

This work focuses on ObsID 1200120197, made on MJD 58305, since this observation covers the transition from type-C to type-B QPOs. To study the source properties during the QPO transition, we split this observation into a total of 15 segments that correspond, more or less, to \textit{NICER} orbits. From now on we call these segments orbits. We discard the last 2 orbits because there is no significant QPO signal while the first 10 orbits show type-C QPOs and the remaining 3 orbits show type-B QPOs (see \citealt{Homan2020} for the QPO identification). The transition time of type-C to type-B QPOs happens on $\sim$MJD 58305.66 \citep[see also][]{Homan2020}.

\subsection{Light curve and hardness-intensity diagram}

To have a global view of the 2018 outburst of MAXI J1820+070, we process the ObsIDs 1200120101$-$1200120312, in total 134 observations (MJD 58190$-$58443). We compute the background-subtracted light curve of the source in the 0.5$-$10\,keV energy band for each ObsID separately. 
We also produce the hardness intensity diagram (HID) with the hardness ratio (HR) defined as the ratio of count rates in the 2.0$-$10.0\,keV band to that in the 0.5$-$2.0\,keV band.

\subsection{Time-averaged spectra}
\label{subsec:method_SSS}

We extract the source and background spectra using \textit{nibackgen3C50}, and generate ARF and RMF files using \textit{nicerarf} and \textit{nicerrmf}, respectively. We group the time-averaged spectra such that there are a minimum of 25 counts per bin and we oversample the intrinsic resolution of the instrument by a factor of 3.

We fit the time-averaged spectra with {\sc xspec} v12.12.1 \citep[][]{Arnaud1996}, using the 0.5$-$10\,keV band data adding a 1\% systematic error to the data below 3\,keV to account for calibration uncertainties\footnote{\url{https://heasarc.gsfc.nasa.gov/docs/nicer/analysis_threads/plot-ratio/}}. First, we jointly fit the time-averaged spectra of all 13 orbits with the model {\tt TBfeo*(diskbb+nthComp)} in {\sc xspec}. {\tt TBfeo} represents the neutral absorption from the interstellar medium in the direction of the source with the abundance and cross-section tables from \citet{Wilms2000} and \citet{Verner1996}, respectively. In the component {\tt TBfeo}, the parameter $N_{\rm H}$ describes the hydrogen column density along the line of sight to the source. This model component also allows to fit the abundance of Fe and O in the absorber. We fixed these abundances to solar in our fits. We use the component {\tt diskbb} \citep{Mitsuda1984, Makishima1986} to model the optically thick and geometrically thin disc \citep{Shakura1973}. This component has two parameters: the inner disc temperature, $kT_{\rm in}$, and a normalization. We use the component {\tt nthComp} \citep{Zdziarski1996, Zycki1999} to model the thermal Comptonization component, with the following parameters: the photon index, $\Gamma$, the corona temperature, $kT_{\rm e}$, the seed photon temperature, $kT_{\rm bb}$, and a normalization. In our fitting, we link $kT_{\rm in}$ in {\tt diskbb} with $kT_{\rm bb}$ in {\tt nthComp} since we assume that the seed photons originate from the disc. 

The fit with this model shows significant residuals at 6--7\,keV, likely due to the reflection component \citep[see also,][]{Wang2021}. Therefore, we add the relativistic reflection component, {\tt relxillCp}\footnote{\url{http://www.sternwarte.uni-erlangen.de/~dauser/research/relxill/}} \citep[][]{Dauser2014}, to the model\footnote{In this paper we use {\tt relxill} version 1.4.3 because the latest version, v2.3, assumes a factor of five lower temperature of the seed-photon source than this one; as we explain in this, this difference has an impact on the low-energy part of the spectrum. In Appendix~\ref{Appendix-A} we discuss the differences in the fits using the latest version of {\tt relxill}.}. We note that the seed photons in {\tt relxillCp} correspond to the Comptonized {\tt nthComp} spectrum of a disk-blackbody source with a fixed temperature of $kT_{\rm in} = 0.05$\,keV, which in the X-ray band is essentially identical to a power-law with a high-energy cutoff. This spectrum is very different from the Comptonized spectrum of low-mass X-ray binaries (LMXBs), with typical disk-blackbody temperatures $kT_{\rm in} \sim$0.5--2\,keV. This choice introduces a bias in the spectral modeling, which manifests as a soft excess at energies $\lesssim$2\,keV that becomes more evident in low-energy {\it NICER} data when the absorption column is low, as in MAXI~J1820+070 ($N_{\rm H} \approx 0.04 \times 10^{22}$\,cm$^{-2}$, see Section~\ref{sec:results_sss}). As a first-order attempt to correct for this soft-excess introduced by {\tt relxillCp}, we renormalize this component by the ratio between an {\tt nthComp} model evaluated at $kT_{\rm in}=kT_{\rm diskbb}$ and the {\tt nthComp} at 0.05~keV, using the so-called {\tt nthratio}\footnote{The {\sc nthratio} model applies an empirical correction to {\tt relxillCp} to account for the fact that the {\tt relxillCp} tables have been computed for a fixed seed-photon temperature of 0.05\,keV, whereas in galactic black-hole binaries the temperatures are $\sim$10 times higher than that. The model is publicly available at \url{https://github.com/garciafederico/nthratio}.} model. We note that {\tt nthratio} adds no extra parameters because all its parameters are linked to the corresponding parameters of {\tt nthComp}. Considering the calibration uncertainties of the data around 0.5\,keV, we also add an extra Gaussian absorption component, {\tt gabs}, to the model, which is now {\tt TBfeo*(diskbb+nthratio*relxillCp+nthComp)*gabs}. We link $kT_{\rm e}$ and $\Gamma$ in {\tt relxillCp} with the corresponding parameters $kT_{\rm e}$ and $\Gamma$ in {\tt nthComp}. We also fix $kT_{\rm e}$ and the BH spin, $a_{\rm *}$, at 40\,keV and 0.998, respectively, consistent with the values given by \citet{Wang2021}. We also link the inclination, $i$, and iron abundance, $A_{\rm Fe}$, in all 13 orbits during the joint fit. As before, we also link the hydrogen column density, $N_{\rm H}$, of the 13 orbits. To get only the reflected emission of {\tt relxillCp}, we fix the reflection fraction to --1. We set the disc inner radius, $R_{\rm in}$, at the ISCO, link the emissivity indices of the {\tt relxillCp}, $\alpha1$, and $\alpha2$, to be the same, and we let the ionization parameter, ${\xi}$, and the normalization of the {\tt relxillCp} component fit freely.

\subsection{Power spectra}
\label{subsec:method_PDS} 

We use {\sc GHATS}\footnote{{\sc ghats} is accessible at \url{http://www.brera.inaf.it/utenti/belloni/GHATS_Package/Home.html}.} to generate PDS in the full energy band (0.5$-$12\,keV). We set the length of the segment and time resolution of each Fast Fourier Transformation (FFT) to, respectively, 16\,s and 0.25\,ms, such that the corresponding lowest and Nyquist frequency of each FFT is 0.06\,Hz and 2000\,Hz, respectively. We average the PDS of all segments to obtain a PDS per orbit. We subtract the Poisson level using the averaged power in the frequency range of \textgreater100\,Hz, and apply a logarithmic rebin in frequency such that the size of each bin is a factor exp(1/100) larger than that of the previous bin. We finally normalize the PDS to units of $\rm{rms}^{2}$ per Hz \citep{Belloni1990} without considering the background count rate because it is negligible compared to the count rate of the source.

We fit the PDS with a combination of Lorentzians \citep{Nowak2000, Belloni2002} in the frequency range 0.06--30\,Hz using {\sc xspec} v12.12.1. For orbits with type-C QPOs the model consists of three Lorentzians for the QPO, high- and low-frequency noise/subharmonic of the QPO, respectively, plus an extra Lorentzian for the second harmonic of the QPO. For orbits with type-B QPOs we fit three Lorentzians for the QPO fundamental and high- and low-frequency noise components. 

To obtain the rms spectra of the source, we compute FFTs in 8 separate sub-bands: 0.5$-$0.75, 0.75$-$1.0, 1.0$-$1.5, 1.5$-$2.5, 2.5$-$4.0, 4.0$-$5.0, 5.0$-$6.5, and 6.5$-$12.0\,keV using the same length of the FFT and Nyquist frequency as for the full-band PDS. We normalized the PDS in each band to rms$^{2}$ per Hz and fitted them with the same multi-Lorentzian model we used fit the full-band PDS. We checked that the centroid frequency and FWHM of all the Lorentzian are consistent with being energy-independent, therefore we take the best-fitting value of these parameters from the full energy band and fix them for the fits to the PDS of the 8 sub-bands, letting only the normalization free during the fit. Finally, we calculate the square root of the normalization of the best-fitting Lorentzianes to compute the fractional rms amplitude and obtain the energy-dependent rms spectra of the QPOs. We also compute the 1-$\sigma$ confidence error of the rms amplitude.

\subsection{Cross spectra}
\label{subsec:method_cross-pha} 

We compute the Fourier cross-spectra of each orbit to calculate the phase lags of the QPO in different energy bands. To do this we use the same time resolution and segment length as for the PDS. The energy bands selected are the same as for the rms spectra, and we use the full energy band (0.5$-$12\,keV) as the reference band. We also compute cross spectra and phase lags of the QPO in two broad energy bands, the reference and subject bands being 0.5--2.0\,keV and 2.0--12.0\,keV, respectively. For the fits of the cross-spectra in the sub-bands, we take the parameters of the best-fitting model of the PDS in the full energy band as the baseline, fixing the centroid frequency and FWHM of each Lorentzian, and letting only the normalization free when we fit jointly the real and imaginary parts of the cross spectra. We added a constant to the model of the real part of the cross spectrum to account for the correlation introduced by the fact that the photons in the subject bands are also in the reference band. The phase lags of the QPO are therefore $\Delta \phi=\arctan (\frac{N_{\rm Im}}{N_{\rm Re}}$), where $N_{\rm Im}$ and $N_{\rm Re}$ are the normalizations of the QPO Lorentzian component in the imaginary and real part of the cross-spectrum, respectively \citep[see][and \citealt{Alabarta2022} for more details]{Peirano2022}. As with the rms amplitude, the uncertainties of the phase lag correspond to the 68\% confidence range.

\subsection{Joint fitting of the time-averaged, rms and phase-lag spectra}
\label{sec:method_jointfit}

Finally, to infer the properties of the corona, we fit the time-averaged spectra of the source plus the rms spectra and phase-lag spectra of the QPOs simultaneously. We need to mention here that, ideally, we should jointly fit the 39 spectra (i.e., 13 time-averaged, 13 rms, and 13 phase-lag spectra) of the orbits with QPOs. However, since there are too many data sets, such joint fitting is very complex and time-consuming. We therefore choose to fit, for each orbit, jointly the time-averaged spectra of the source and the rms and lag spectra of the QPOs, fixing $N_{\rm H}$ and the source inclination and iron abundance in {\tt relxillCp} to the values we obtained from the simultaneous fit to the 13 time-averaged spectra (see Section~\ref{subsec:method_SSS}). Other parameters in the components {\tt diskbb}, {\tt relxillCp}, and {\tt nthComp} are free to vary \citep[see also][]{Zhang2022}, except that $kT_{\rm e}$ is fixed at 40\,keV. 

Before we explain the details of the joint fitting, we give a description of the time-dependent Comptonization model {\tt vkompth} and its parameters. The steady-state version of the {\tt vkompth} model is the same as the thermal Comptonized model {\tt nthComp}, whereas the model has extra parameters that only affect the time-dependent version of the model. There are two versions of the {\tt vkompth} model, for one or two coronae. For the model of one corona, called, {\tt vkompthdk} or {\tt vkompthbb} depending on whether the seed-photon source is, respectively, a disc or a blackbody, the model parameters are the electron temperature, $kT_{\rm e}$, the seed photons temperature, $kT_{\rm s}$, the photon index, $\Gamma$, the corona size, $L$, the feedback fraction, $\eta$, the amplitude of the variability of the external heating rate, $\delta \Dot{H}_{\rm ext}$, the size of the seed-photon source, $a_{f}$, and the reference lag, \textit{reflag}. The model for two different, but physically connected, coronae are called {\tt vkdualdk} and {\tt vkdualbb} \citep{Garcia2021, Bellavita2022}. For the \textit{dual} corona models, there are two sets of parameters $kT_{\rm s,1}/kT_{\rm s,2}$, $kT_{\rm e,1}/kT_{\rm e,2}$, $\Gamma_{1}/\Gamma_{2}$, $L_{1}/L_{2}$, $\eta_{1}/\eta_{2}$, and $\delta \Dot{H}_{\rm ext,1}/\delta \Dot{H}_{\rm ext,2}$ to describe the physical properties of the \textit{small(1)} and \textit{large(2)} coronae, respectively. The remaining parameters of the \textit{dual} models are the size of the seed-photon source and the reference lag, plus an additional parameter, $\phi$, that describes the phase difference between the two coronae \citep[for more details see][]{Garcia2021}.

First, we perform a joint fit using the model for one corona and seed photons from a disc, {\tt vkompthdk}. Although the model generally reproduces the data, the fits give relatively large $\chi^2$, with the largest residuals coming from the rms and lag spectra. We, therefore, replace {\tt vkompthdk} by {\tt vkdualdk} with two Comptonized regions. The reduced chi-square decreases from 288.69 for 234 degrees of freedom (dof) for one corona to 230.89 for 230 dof for two coronae for a typical data set. We note that the $\chi^2$ and dof are dominated by the time-averaged spectrum since this one has the largest number of energy channels. The improvement in the $\chi^2$ when going from a model with one to a model with two coronae happens in the fits to the rms and lag spectra of the QPOs. For instance, in the example mentioned above the $\chi^2$ of the fit to the rms and lag spectra of the QPO goes, respectively, from 33.18 to 20.87 and from 51.41 to 6.49 for 8 channels, showing that a model with two coronae improves the fit significantly compared to that with only one corona.

We therefore fit together the time-averaged spectra of the source and the rms and lag spectra of the QPO of all orbits as follows: (i) We use the model described in Section~\ref{subsec:method_SSS} to fit the time-averaged spectra of the source, with the parameters setting being the same as in Section~\ref{subsec:method_SSS}. (ii) We fit the rms spectra of the QPO with the model {\tt vkdualdk*dilution}, where the component {\tt dilution} is the ratio of the energy-dependent flux density of the Comptonized component to the flux density of the total continuum, and is a factor that accounts for the decrease of the observed rms amplitude compared to the amplitude of the corona given by the model due to the emission of the non-variable components in the energy spectrum of the source \citep[see][]{Bellavita2022}. The {\tt dilution} component is necessary because we assume that the QPO originates from the corona, and the other components (such as {\tt diskbb} and {\tt relxillCp}) do not contribute, or at least not at the QPO frequency, to the variability. We link the parameters $kT_{\rm s,1}$, $kT_{\rm e,1}=kT_{\rm e,2}$ and $\Gamma_{1}=\Gamma_{2}$ in {\tt vkdualdk} to the corresponding parameters $kT_{\rm in}$, $kT_{\rm e}$ and $\Gamma$ in the {\tt diskbb} and {\tt nthComp} components that fit the time-averaged spectra, and we let the parameters $kT_{\rm s,2}$, $L_{1}$, $L_{2}$, $\eta_{1}$, $\eta_{2}$, $\delta \Dot{H}_{\rm ext,1}$, $\delta \Dot{H}_{\rm ext,2}$, $\phi$, and $reflag$ free during the fit. We link all the parameters of the {\tt dilution} component to the corresponding parameters of the model of the time-averaged spectrum. We note that the {\tt vkompth} model is not sensitive to the size of the seed-photon emitting region, $a_{f}$, therefore we fix $a_{f}$ to 250\,km in our work \citep[for more details see][]{Garcia2022}. (iii) For the phase-lag spectra of the QPO, the model is similar to the model for the rms spectra but without the dilution component, i.e., the fitting model is {\tt vkdualdk}. The parameters of the {\tt vkdualdk} model for the phase-lag spectra are all linked to the same parameters for the model of the rms spectra.

\section{Results}
\label{sec:results}

\begin{figure*}
\centering
\includegraphics[width=\textwidth]{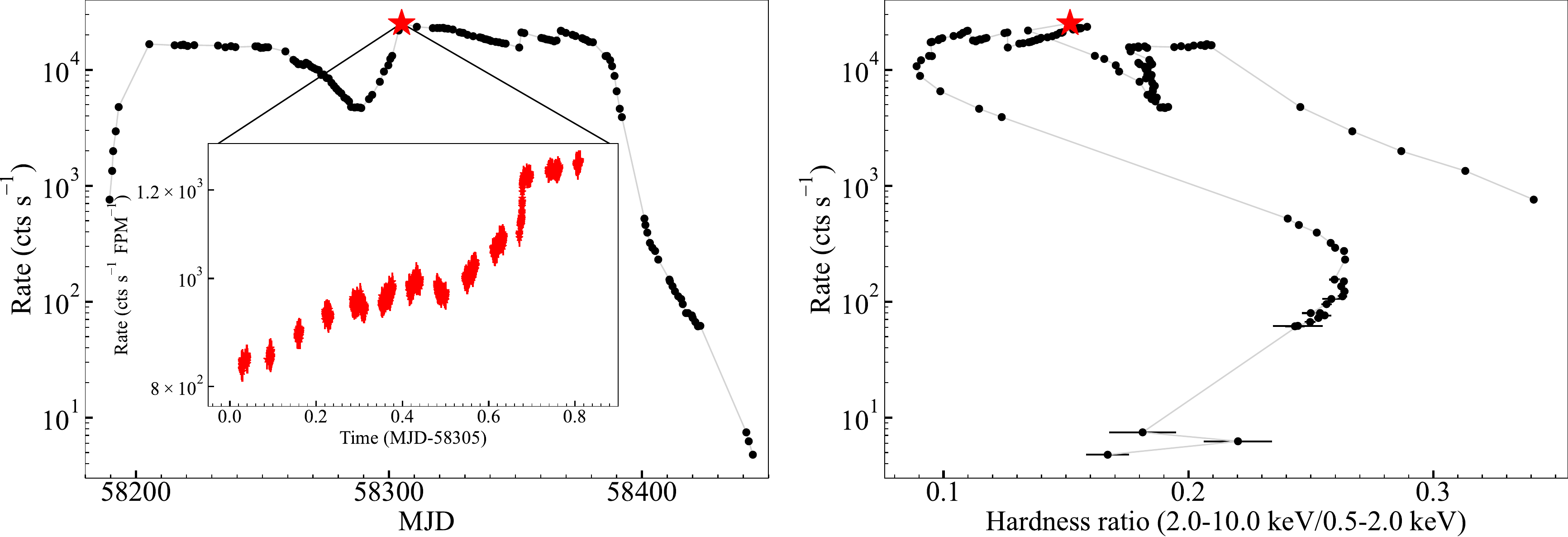} \\ 
\caption{\textit{NICER} light curve (left panel) and hardness-intensity diagram (right panel) of MAXI J1820+070 during the 2018 outburst. In both panels, the intensity is the background-subtracted count rate in the 0.5$-$10\,keV band, while the hardness ratio in the right panel is the ratio of the count rate in the 2$-$10\,keV band to that in the 0.5$-$2\,keV band. Each point corresponds to one \textit{NICER} ObsID. The observation used in our work is marked in red. In the left panel, the inset shows the background-subtracted light curve (0.5--10~keV) of the observation used in this paper, with a time resolution of 10~s.}
\label{fig:LC_HID}
\end{figure*}

\subsection{Light curve and HID}

We present the 0.5$-$10\,keV \textit{NICER} light curve of MAXI J1820+070 during the 2018 outburst with one point per obsID on the left panel of Figure~\ref{fig:LC_HID}. At the beginning of the outburst the count rate increases rapidly from $\sim$760 cts\,${\rm s}^{-1}$ on MJD 58196 to $\sim$16600 cts\,${\rm s}^{-1}$ on MJD 58205. The count rate then remains more or less constant except for a small excursion down and up between MJD 58264--58304 that reaches a count rate as low as $\sim$4700 cts\,${\rm s}^{-1}$. From MJD 58304--58311, the source is in the intermediate state, and then stays in the soft state until MJD 58380 \citep[e.g.,][]{Shidatsu2019}. After MJD 58381, the count rate drops rapidly from $\sim$17000 cts\,${\rm s}^{-1}$ to $\sim$5 cts\,${\rm s}^{-1}$ on MJD 58443, and after that, the source is almost undetectable. The observation on MJD 58305 that we use in our work is marked in red. In that observation, the source is in the intermediate state, with the QPO transitioning from type-C to type-B \citep{Homan2020}. The inset in the left panel of Figure~\ref{fig:LC_HID} shows the light curve of this observation with a time resolution of 10 s.

We also display the HID of the source on the right panel of Figure~\ref{fig:LC_HID}. The trace of the HID shows a typical ``q'' shape seen in most BHTs \citep[e.g.,][]{Fender2004, Belloni2016}. In the early stage of the outburst, the count rate of MAXI J1820+070 increases rapidly as the HR decreases from 0.3 to 0.2, and the source moves rapidly from the right branch of the HID to the upper branch. In the upper branch, the HR continues decreasing to 0.09 while the count rate remains more or less constant. The count rate then decreases rapidly, with an increase of the HR to 0.26, indicating that the source reaches the end of the outburst. The observation used in this work is marked in red in this panel.

\subsection{Time-averaged spectra results}
\label{sec:results_sss}

We fit the time-averaged spectra of the 13 orbits simultaneously. As described in Section~\ref{subsec:method_SSS}, we link hydrogen column density, $N_{\rm H}$, inclination angle, $i$, and iron abundance, $A_{\rm Fe}$, to be the same for the 13 orbits. Our fitting results give a relatively low $N_{\rm H}=4.17^{+0.25}_{-0.15}\times10^{20}\,\rm{cm}^{-2}$, consistent with $N_{\rm H} \sim 5\times10^{20}\,\rm{cm}^{-2}$ given by \cite{Fabian2020}. The inclination angle $i={66.9^{\circ}}^{+0.2}_{-1.0}$, is consistent with previous reports of the orbital inclination \citep[$66^{\circ} < i < 89^{\circ}$,][]{Torres2020} and the jet angle ($i=63 \pm 3^{\circ}$, \citealt{Atri2020}; $i=64 \pm 5^{\circ}$, \citealt{Wood2021}). The chi-square of the joint fitting is 3036.7 for 2932 dof.

\begin{figure*}
\centering
\includegraphics[width=\textwidth]{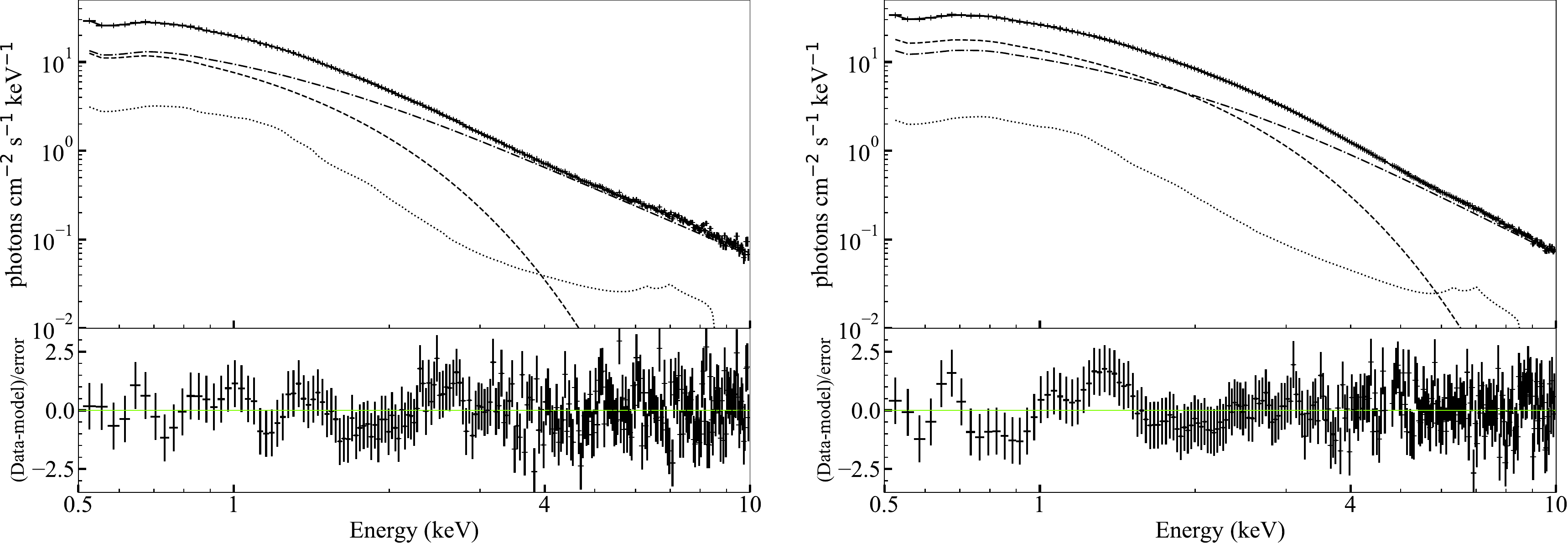} \\ 
\caption{Representative time-averaged spectra of MAXI J1820+070 in the 0.5$-$10\,keV energy band with \textit{NICER}. Left panel: Time-averaged spectrum of orbit 01 with a type-C QPO. Right panel: Time-averaged spectrum of orbit 10 with a type-B QPO. In both panels, the upper plot shows the data, the total model (black line), and the {\tt diskbb} (dashed line), {\tt nthComp} (dashed-dotted line), and {\tt relxillCp} (dotted line) components, respectively, while the lower panels show the residuals of the best-fitting model. 
\label{fig:ss_pha}}
\end{figure*}

We display two representative 0.5--10.0\,keV time-averaged spectra of an orbit with a type-C and another with a type-B QPO in Figure~\ref{fig:ss_pha}. The evolution of the geometrical and spectral parameters will be described in Section~\ref{sec:results_jointfit}.

\begin{figure*}
\centering
\includegraphics[width=\textwidth]{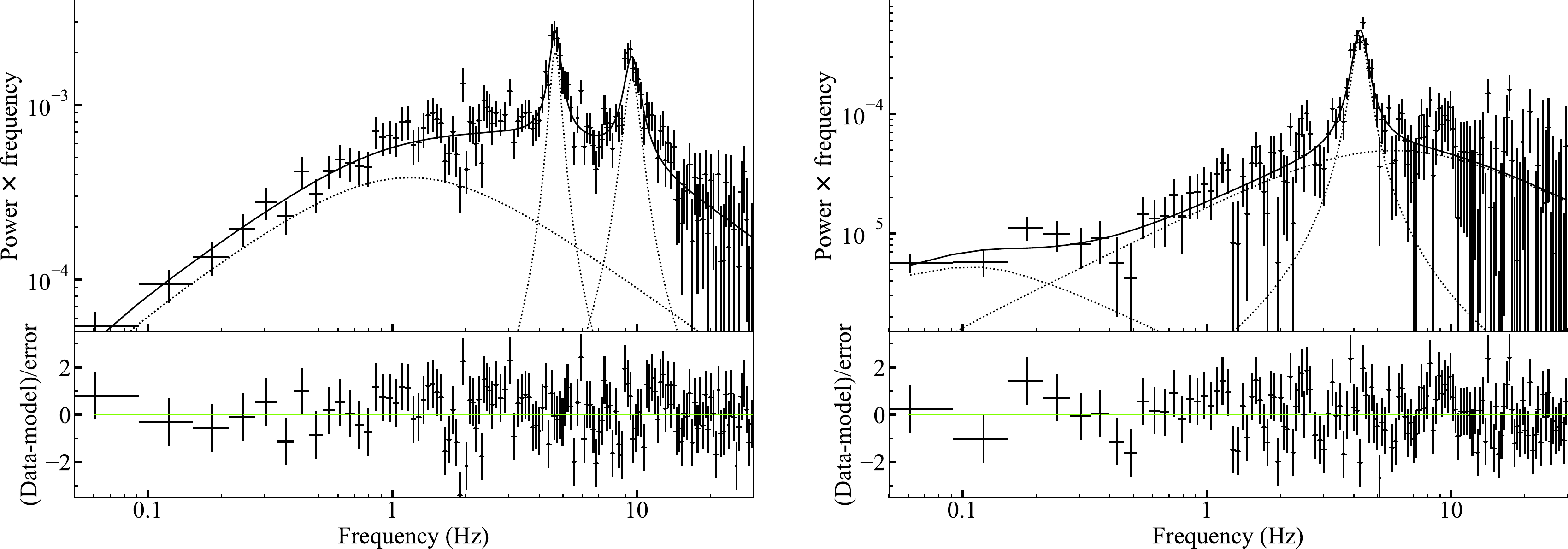} \\ 
\caption{Representative PDS of MAXI J1820+070 in the 0.5$-$12\,keV energy band with \textit{NICER}. Left panel: PDS of orbit 01 with a type-C QPO. Right panel: PDS of orbit 10 with a type-B QPO. In both panels, the upper plot shows the data, the total model (black line), and the individual Lorentzian components (dotted lines), while the lower panels show the residuals of the best-fitting model. }
\label{fig:PDS}
\end{figure*}

\begin{table*}
    \renewcommand\arraystretch{1.4}
	\centering
	\caption{Observation log and parameters of the QPOs of MAXI J1820+070. Uncertainties are given at the 90\% confidence level.}
	\label{tab:freq_evo}
	\begin{threeparttable}
	\resizebox{\linewidth}{!}{
	\begin{tabular}{ccccllllc} 
		\hline
		orbit & Start time & Stop time & Rate$^{a}$ & frequency & FWHM & full-band rms$^{b}$ & average phase lag$^{c}$ & QPO type\\
		 & (MJD--58305) & (MJD--58305) & (cts~${\rm s}^{-1}$) & (Hz) & (Hz) & (\%) & (rad) & \\
		\hline
		00 & 0.02704 & 0.04084 & $20800\pm{30}$ & $4.44\pm{0.07}$ & $0.5^{+0.6}_{-0.2}$ & $1.7^{+0.5}_{-0.2}$ & $0.17\pm{0.13}$ & C\\	
		01 & 0.09140 & 0.09656 & $21040\pm{30}$ & $4.64\pm{0.05}$ & $0.52^{+0.18}_{-0.13}$ & $1.89^{+0.23}_{-0.18}$ & $0.07\pm{0.09}$ & C\\
		02 & 0.15593 & 0.16398 & $22060\pm{30}$ & $5.01\pm{0.05}$ & $0.58^{+0.30}_{-0.18}$ & $1.56^{+0.19}_{-0.15}$ & $0.06\pm{0.11}$ & C\\
		03 & 0.22011 & 0.23366 & $22970\pm{30}$ & $5.50\pm{0.05}$ & $0.54^{+0.15}_{-0.12}$ & $1.29\pm{0.11}$ & $0.12\pm{0.11}$ & C\\
		04 & 0.28464 & 0.31375 & $23500\pm{30}$ & $5.83\pm{0.05}$ & $0.64^{+0.19}_{-0.16}$ & $1.08\pm{0.10}$ & $-0.09\pm{0.09}$ & C\\
		05 & 0.35088 & 0.37810 & $23940\pm{30}$ & $6.04\pm{0.06}$ & $1.0\pm{0.2}$ & $1.23\pm{0.11}$ & $-0.16\pm{0.10}$ & C\\
		06 & 0.41478 & 0.44108 & $24620\pm{30}$ & $6.56\pm{0.11}$ & $1.5\pm{0.6}$ & $1.22^{+0.20}_{-0.16}$ & $-0.22\pm{0.13}$ & C\\
		07 & 0.47925 & 0.50079 & $24190\pm{30}$ & $6.34\pm{0.05}$ & $0.49^{+0.18}_{-0.13}$ & $0.94\pm{0.10}$ & $-0.27\pm{0.15}$ & C\\
		08 & 0.54464 & 0.57048 & $25240\pm{30}$ & $7.2\pm{0.2}$ & $2.7\pm{0.9}$ & $1.22\pm{0.19}$ & $-0.34\pm{0.13}$ & C\\
		09 & 0.60972 & 0.63542 & $26800\pm{30}$ & $7.7\pm{0.3}$ & $0.81^{+1.6}_{-0.6}$ & $0.47^{+0.28}_{-0.14}$ & $0.8^{-1.2}_{+0.5}$ & C\\
  \hline
		10 & 0.67056 & 0.69663 & $30220\pm{40}$ & $4.23\pm{0.04}$ & $0.61\pm{0.09}$ & $1.01\pm{0.06}$ & $-0.73\pm{0.10}$ & B\\
		11 & 0.73783 & 0.76422 & $31490\pm{40}$ & $3.98\pm{0.04}$ & $0.59\pm{0.11}$ & $0.77\pm{0.05}$ & $-0.78\pm{0.10}$ & B\\
		12 & 0.80227 & 0.81218 & $23650\pm{30}$ & $3.14\pm{0.17}$ & $0.6^{+0.8}_{-0.3}$ & $0.48^{+0.15}_{-0.12}$ & $-1.2\pm{0.6}$ & B\\
		\hline
	\end{tabular}
	}
    \begin{tablenotes} 
    \footnotesize{
    \item[$^{a}$] Count rate of the source in the 0.5--10.0\,keV band.
    \item[$^{b}$] The rms amplitude of the QPO in the 0.5--12.0\,keV band.
    \item[$^{c}$] The phase lag of the QPO between photons in the 2.0--12.0\,keV band with respect to those in the 0.5--2.0 keV band.}
    \end{tablenotes} 
    \end{threeparttable}
\end{table*}

\begin{figure*}
\centering
\includegraphics[width=\textwidth]{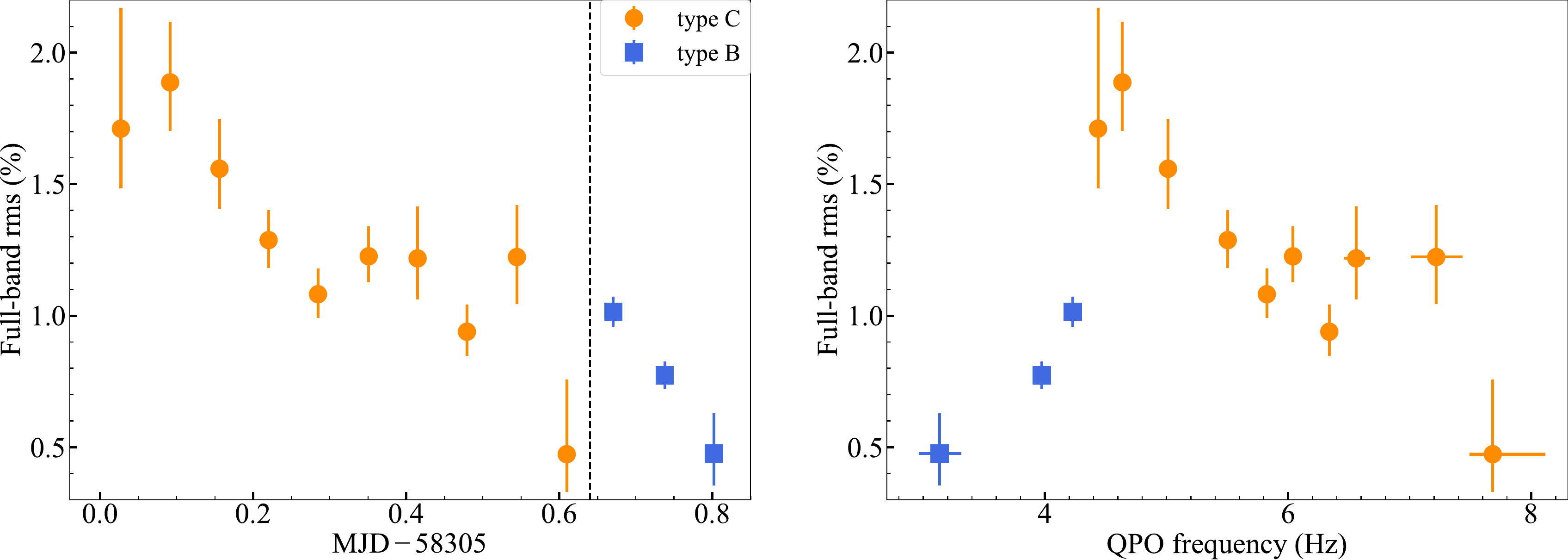} \\ 
\caption{Left panel: Evolution of full-band (0.5--12\,keV) rms amplitude of the QPO during the transition from type-C to type-B QPOs for MAXI J1820+070. The vertical dashed line represents the transition time from type-C to type-B QPO. Right panel: Full-band rms amplitude of the QPO as a function of the centroid frequency of QPO. In both panels, orange and blue points represent orbits with type-C and type-B QPOs, respectively. \\
\label{fig:Rms_evo}}
\end{figure*}


\subsection{Timing results}

\subsubsection{rms and phase lags of the QPOs as a function of time}
\label{subsec:timing_evo}

We present representative 0.5--12\,keV PDS with the type-C and type-B QPOs in Figure~\ref{fig:PDS}. There is a significant second harmonic in the case of the type-C but not for the type-B QPOs. The centroid frequency of the type-C QPOs increases gradually with time during the observation, from 4.4\,Hz to 7.7\,Hz, while that of the type-B QPOs decreases gradually from 4.2\,Hz to 3.1\,Hz (see Table~\ref{tab:freq_evo}). We need to mention that during the transition from type-C to type-B QPOs in orbit 09, due to the rapid decrease of the significance of the type-C QPO, the errors of the parameters are large in the data. 

We show the evolution of the full-band rms amplitude of the QPOs of the 13 orbits in the 0.5$-$12.0\,keV energy band on Figure~\ref{fig:Rms_evo}. The full-band rms amplitude of the type-C QPO gradually decreases from $\sim$2\% to $\sim$1\% with time and then remains more or less constant, whereas the full-band rms amplitude of the type-B QPO is lower than that of the type-C QPO and decreases gradually from $\sim$1\% to $\sim$0.5\%. As shown on the right panel of Figure~\ref{fig:Rms_evo}, the full-band rms amplitude of type-C and type-B QPOs show different correlations as a function of the centroid frequency of the QPOs. The average rms amplitude of type-B QPOs appears to be positively correlated with the QPO centroid frequency (although there are only three measurements of the type-B QPO), whereas there is an anti-correlation in the case of type-C QPOs. We note that the average rms amplitude of the last type-C QPO point is relatively small ($\sim$0.5\%), which is probably due to the weaker type-C QPO signal and the rapid change of the QPO frequency during the transition from type-C to type-B QPOs.

\begin{figure*}
\centering
\includegraphics[width=\textwidth]{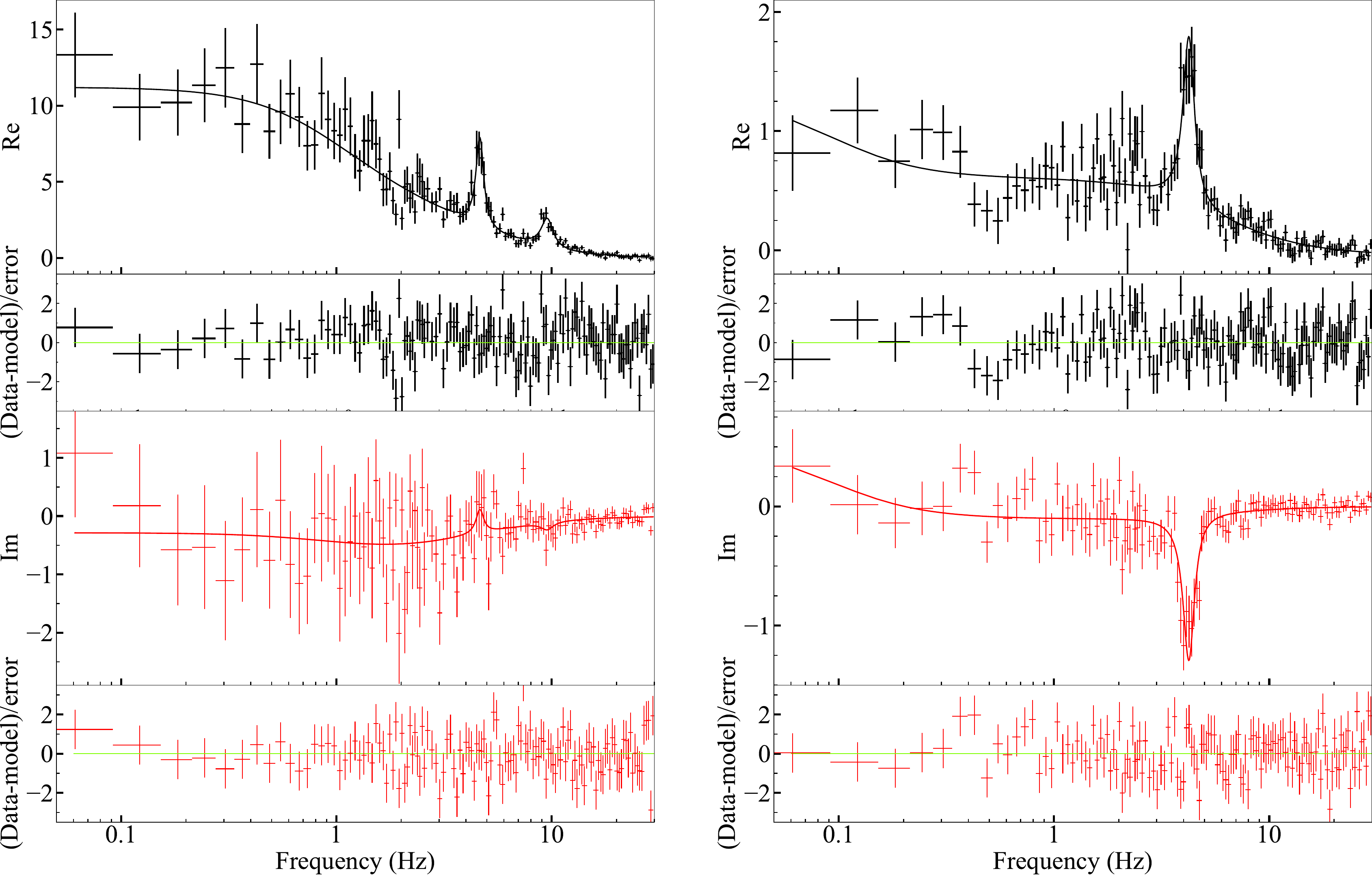} \\ 
\caption{Representative cross-spectra of the $2.0-12.0$ keV band with respect to the $0.5-2.0$ keV band of MAXI J1820+070. The black and red symbols and lines represent, respectively, the real and imaginary parts of the cross-spectrum. Left panel: Cross spectrum of orbit 01 with a type-C QPO. Right panel: Cross spectrum of orbit 10 with a type-B QPO. In both panels, we show the best-fitting model to the real and imaginary part of the cross-spectrum and the residuals of the best-fitting model.
\label{fig:Cross_spectra}}
\end{figure*}

We present representative cross spectra ($2.0-12.0$ keV with respect to $0.5-2.0$ keV) of one type-C and one type-B QPOs in Figure~\ref{fig:Cross_spectra}. The shape of the cross spectra when the type-C and type-B QPOs are at the same centroid frequency is different. The real part of the cross-spectrum at the QPO frequency is positive for both QPOs, but the real part of the type-C QPO is larger than that of the type-B QPO. On the other hand, the imaginary part of the cross-spectrum of the type-C QPO is small and positive, whereas for the type-B QPO it is large and negative. This indicates that in these examples, when comparing these two energy bands, the type-C QPO has a small positive phase lag whereas the type-B QPO has a relatively large negative phase lag (see Section~\ref{subsec:method_cross-pha} for details on the calculation of the phase lags).

\begin{figure*}
\centering
\includegraphics[width=\textwidth]{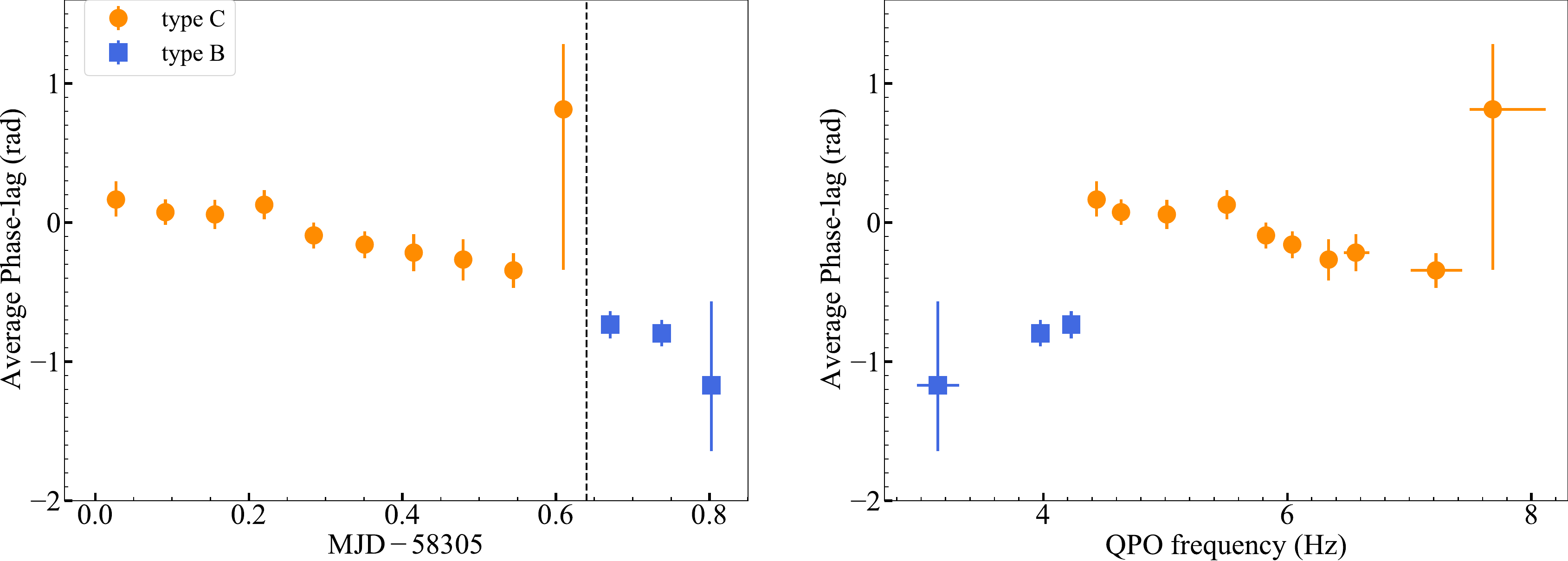} \\ 
\caption{Left panel: Evolution of the QPO phase-lags of the $2.0-12.0$\,keV band with respect to $0.5-2.0$ keV band of MAXI J1820+070 during the transition from type-C to type-B QPOs. The vertical dashed line represents the transition time from type-C to type-B QPO. Right panel: QPOs phase-lags as a function of the centroid frequency of the QPO. Orange and blue points represent orbits with the type-C and type-B QPOs, respectively. The large error bars associated with orbit 09 (last orange point) are due to the lower significance of the type-C QPO in that segment, possibly because the transition has already started.} 
\label{fig:Lag_evo}
\end{figure*}

We show the evolution of the average phase lags of the QPOs on the left panel of Figure~\ref{fig:Lag_evo}. The average phase lags of both type-C and type-B QPOs gradually decrease with time, from $\sim$0.2\,rad to $\sim-$0.3\,rad and from $\sim-$0.7\,rad to $\sim-$1.2\,rad, respectively. We also show the dependence of the average phase lags on the centroid frequency of the QPO on the right panel of Figure~\ref{fig:Lag_evo}. In the case of type-C QPOs, the average phase lags are anti-correlated with the QPO frequency, whereas for the type-B QPOs the average phase lags are either positively or uncorrelated with the centroid frequency. Again, for the last point of type-C QPO, the average phase lag has a small value with a large error bar due to the lower significance of the QPO.

\begin{figure*}
\centering
\includegraphics[width=\textwidth]{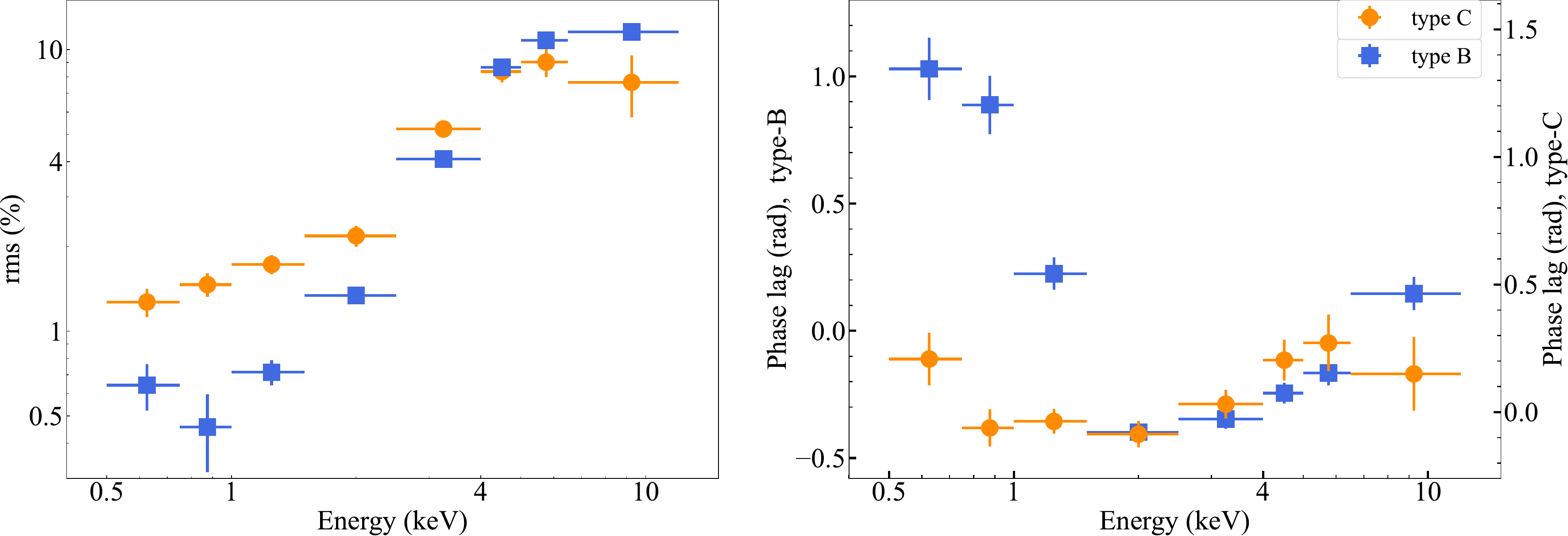} \\ 
\caption{Representative rms spectra (left panel) and phase-lag spectra (right panel) of the type-C and type-B QPO of MAXI J1820+070. The orbits are the same as in Figure~\ref{fig:PDS}. 
\label{fig:RMS-LAG_pha}}
\end{figure*}


\subsubsection{rms and phase-lag spectra of the QPOs}
\label{timing_pha}

To further explore the rms and phase lag energy dependence of different types of QPOs, especially during the QPO transition period, we plot representative rms-energy spectra of the type-B and type-C QPOs in the left panel of Figure~\ref{fig:RMS-LAG_pha}. Both for the type-C and type-B QPOs the rms amplitude increases as the energy increases, similar to the results of several low- and high-frequency QPOs reported in other sources \citep[e.g., ][and references therein]{Sobolewska2006, Mendez2013}. The rms amplitude of the type-C QPOs increases monotonically with energy from 1\% to 10\%. For the type-B QPOs the rms amplitude remains more or less constant at $\sim$$0.5-0.7$\% from 0.5 keV to $\sim$1.5--2~keV, and then increases to a similar level, $\sim$10\%, as for the type-C QPOs at the highest energy band.

We show the corresponding lag-energy spectra of the type-C and type-B QPOs in the right panel of Figure~\ref{fig:RMS-LAG_pha} using the full band as the reference band. For the type-C QPO the phase lags decrease from $\sim-$0.1\,rad to $\sim-$0.4\,rad as the energy increases from 0.5\,keV to 2\,keV, and the phase lags increase again to $\sim-$0.1\,rad when the energy increases up to 12\,keV. For the type-B QPO the phase lags decrease from $\sim$1.0\,rad at the lowest energy band to $\sim-$0.4\,rad as the energy increases from 0.5\,keV to 2\,keV, and then the phase lags increase again to $\sim$0.1\,rad as the energy increases further up to 12\,keV. (Notice that, since the lags are relative quantities, not the actual value of the lags but only the shape of the lag spectrum is important). As it is apparent from both panels of Figure~\ref{fig:RMS-LAG_pha}, the rms and lag spectra of the type-C and type-B QPO are consistent with being the same for energies above $\sim$1.5--2~keV, but as the energy decreases below $\sim$1.5--2~keV the rms spectrum decreases more rapidly and reaches a lower rms amplitude, whereas the lag spectrum increases more rapidly and reaches a larger phase lag value at the lowest energies for type-B than for the type-C QPOs. In fact, from this Figure, it appears that going from type-C to type-B QPOs, the only changes in the rms and lag spectra happen at energies below $\sim$1.5--2~keV.

\begin{figure*}
\centering
\includegraphics[width=\textwidth]{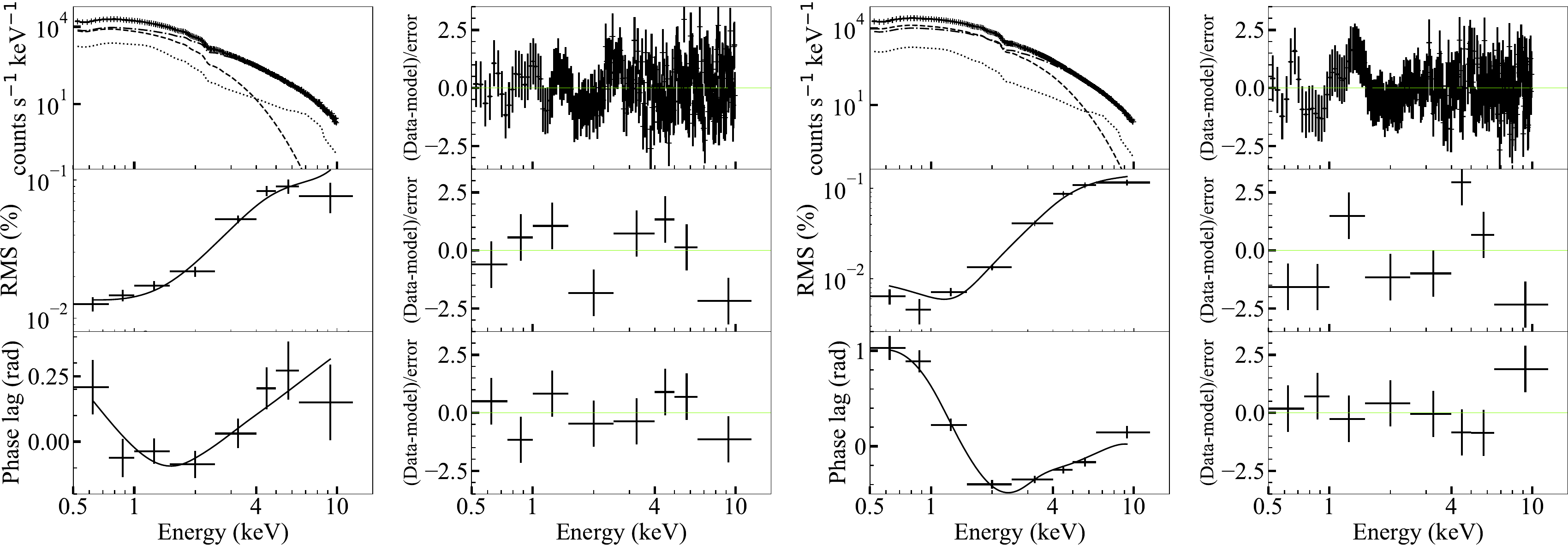} \\ 
\caption{First and third columns: From top to bottom, representative joint fitting of the time-averaged spectrum of the source and the rms and lag spectra of the type-C (first column) and type-B QPO (third column) of MAXI J1820+070. The second and fourth columns show the residuals of the best-fitting model. The orbits for the type-C and type-B QPOs are the same as in Figure~\ref{fig:PDS}.
\label{fig:Joint_fit}}
\end{figure*}

\begin{figure*}
\centering
\includegraphics[width=\textwidth]{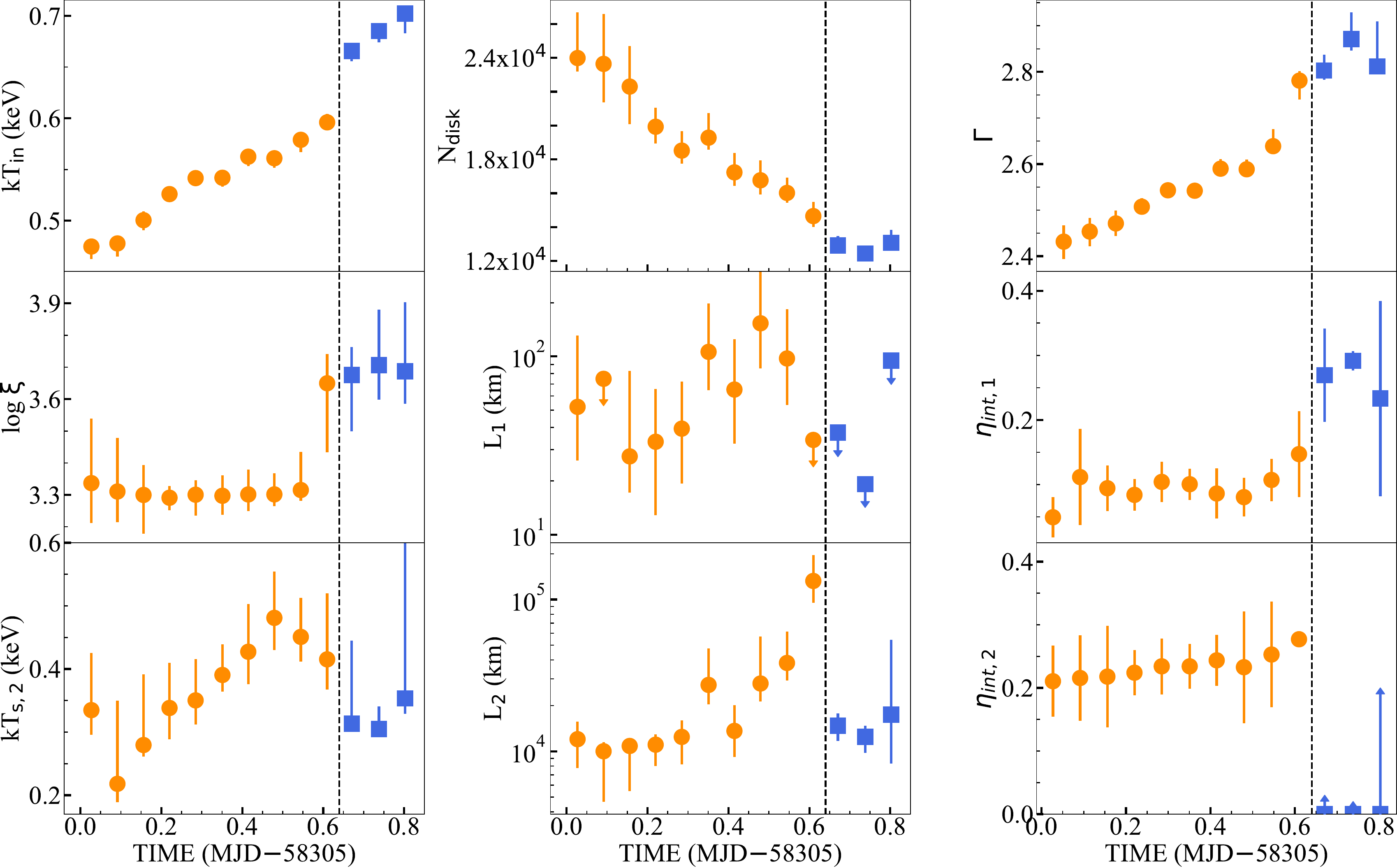} \\ 
\caption{Evolution of the best-fitting geometrical and spectral parameters during the transition from type-C to type-B QPOs of MAXI J1820+070. The parameters are $kT_{\rm in}$, the inner disc temperature, which is linked to the seed photons temperature of the small corona, $kT_{\rm s,1}$, $N_{\rm disk}$, the normalization of {\tt diskbb}, $\Gamma$, the photon index, ${\rm log\,\xi}$, the disc ionization parameter, $L_{1}$/$L_{2}$ and $\eta_{\rm{int,1}}$/$\eta_{\rm{int,2}}$, respectively, the size and the intrinsic feedback fraction of the small/large corona, and $kT_{\rm s,2}$, the temperature of the seed-photon source of the large corona. Uncertainties are at the 90\% confidence level for each parameter. Orange and blue points represent, respectively, orbits with the type-C and type-B QPOs. The vertical dashed line represents the transition time from type-C to type-B QPO. 
\label{fig:Pars_evo}}
\end{figure*}

\begin{table}
    \begin{center}
    \renewcommand\arraystretch{1.4}
	\caption{Representative best-fitting spectral and corona parameters for observations with a type-C and a type-B QPOs in MAXI J1820+070. Uncertainties are given at the 90\% confidence level. All model parameters for all observations are given in Appendix~\ref{tab:Fit_pars_all}.}
	\label{tab:Fit_pars}
	\resizebox{\columnwidth}{!}{
	\begin{tabular}{lccc} 
		\hline
		Component & Parameter & {type-C QPO}$^{\rm (1)}$ & {type-B QPO}$^{\rm (2)}$\\
		\hline
		TBfeo& $N_{\rm H} (10^{20}\,\rm{cm^{-2}})$ & [4.2]$^{a}$ & [4.2]$^{a}$ \\
		diskbb & $kT_{\rm in}$ (keV) & $0.478^{+0.007}_{-0.013}$ & $0.666^{+0.003}_{-0.010}$\\
	    & $N_{\rm disk}\,(10^{4})$ & $2.4\pm{0.3}$ & $1.29^{+0.06}_{-0.03}$\\
		nthComp & $\Gamma$ & $2.45\pm{0.03}$ & $2.80\pm{0.03}$\\
		 & $kT_{\rm e}$ (keV) & [40]$^{a}$ & [40]$^{a}$ \\	
		 & $N_{\rm nthComp}$ & $10.2\pm{0.4}$ & $11.7^{+0.8}_{-0.2}$\\
		relxillCp & $\alpha1$ & $2.23^{+0.20}_{-0.14}$ & $2.06^{+0.32}_{-0.13}$ \\
		 & $a_{\rm *}$ & [0.998]$^{a}$ & [0.998]$^{a}$ \\
		 & $i\,(^{\circ})$ & [67.3]$^{a}$ & [67.3]$^{a}$\\	
		 & ${\rm log}\,\xi\,(\rm{log[erg\,cm\,s^{-1}}])$ & $3.31^{+0.17}_{-0.10}$ & $3.68^{+0.09}_{-0.18}$\\
		 & $A_{\rm Fe}$ (solar) & [10]$^{a,b}$ & [10]$^{a,b}$\\
		 & $N_{\rm refl}$ & $0.10\pm{0.03}$ & $0.19\pm{0.05}$\\
		gabs & $lineE$ (keV) & $0.552^{+0.008}_{-0.021}$ & $0.541^{+0.007}_{-0.033}$\\
		 & $\sigma\,{\rm (10^{-2}\,keV)}$ & $7.8^{+2.0}_{-0.9}$ & $9.9^{+0.7}_{-3.2}$\\
		 & Strength $(10^{-2})$ & $4.9^{+1.4}_{-0.5}$ & $6.9^{+2.4}_{-0.3}$\\
		vkdualdk & $kT_{\rm s,2}$ (keV) & $0.22^{+0.13}_{-0.03}$ & $0.313^{+0.132}_{-0.007}$\\
		& $L_{1}$ (km) & $<74$ & $<36$\\
		& $L_{2}\,{\rm (10^{4}\,km)}$ & $1.0^{+1.5}_{-0.5}$ & $1.5\pm{0.3}$\\
		& $\eta_{1}$ & $0.51^{+0.14}_{-0.40}$ & $0.96^{+0.03}_{-0.39}$\\
		& $\eta_{2}$ & $1.0^{c}_{-0.5}$ & ${{0}^{+0.16}_{d}}$\\
		& $\phi$ (rad) & $2.6^{+1.4}_{-0.4}$ & $3.8^{+0.70}_{-0.14}$\\
		& $reflag$ & $0.07\pm{0.07}$ & $-0.31^{+0.02}_{-0.09}$\\	
		& $\eta_{\rm{int,1}}$ & $0.11\pm{0.07}$ & $0.27\pm{0.07}$ \\
		& $\eta_{\rm{int,2}}$ & $0.22\pm{0.07}$ & ${{0}^{+0.01}_{d}}$ \\
		\hline
		& $\chi^{2}/v$ & 246.02/230 & 233.46/230\\
		& frequency (Hz) & $4.64\pm{0.05}$ & $4.23\pm{0.04}$ \\
		\hline
	\end{tabular}
	}
	\end{center}
\footnotesize{$^{(1)}$ orbit 01.\newline
$^{(2)}$ orbit 10.\newline
$^{a}$ The parameters $N_{\rm H}$, $kT_{\rm e}$, $a_{\rm *}$, $i$, and $A_{\rm Fe}$ have been fixed during the fitting at, respectively, $4.2\times10^{20}\,\rm{cm^{-2}}$, 40\,keV (due to the energy range of the \textit{NICER} data this parameter cannot be constrained), 0.998, 67.3$^{\circ}$, and 10 times solar, as obtained from the joint fits of the energy spectra described in Section~\ref{sec:results_sss}. \newline 
$^{b}$ While in this version of {\tt rexillCp} the iron abundance pegs at the upper limit, if we use the latest version with variable disc density, the iron abundance decreases to approximately 5; the other parameters of the fit remain unchanged.  \newline
$^{c}$ The positive error of the parameter pegged at the upper limit \newline
$^{d}$ The negative error of the parameter pegged at the lower limit.
}	
\end{table}

\subsection{Joint fitting of the time-averaged spectrum of the source and the rms and phase-lag spectra of the QPOs}
\label{sec:results_jointfit}

We show two examples of the best-fitting model and residuals of the data with type-C and type-B QPOs in Figure~\ref{fig:Joint_fit}. We give the best-fitting parameters and the corresponding contour plots for the fits of these two examples in Table~\ref{tab:Fit_pars} and Figures~\ref{fig:Cont1} and ~\ref{fig:Cont10}, respectively (for clarity reasons we only show the contour plots of selected parameters in these Figures, but Table~\ref{tab:Fit_pars} contains the values and errors of all the parameters). We compute the error of the parameters using MCMC, and we give the parameters value and corresponding errors for the 13 orbits in Table~\ref{tab:Fit_pars_all}. For the MCMC simulation, we use the Goodman-Weare algorithm \citep{Goodman2010} with 160 walkers and a total of 160000 samples using a Cauchy proposal distribution after we discard the first 320000 steps of the MCMC to ensure convergence.

\subsection{Evolution of the parameters}
\label{sec:pars_evo_time}

As shown in Figure~\ref{fig:Pars_evo}, for the orbits with type-C QPOs the disc temperature, $kT_{\rm in}$, increases gradually from $\sim$0.5\,keV to $\sim$0.6\,keV while the normalization of the {\tt diskbb} component decreases gradually  from $2.4\times10^{4}$ to $1.5\times10^{4}$. At the same time the photon index, $\Gamma$, increases from 2.4 to 2.7. Compared to the orbits of type-C QPOs, for the orbits with type-B QPOs the Comptonized component is steeper with a photon index $\Gamma\sim$2.8, while the inner disc temperature is higher,  $kT_{\rm in}\sim$0.7\,keV, with a smaller disc normalization $\sim1.3\times10^{4}$. The value of the ionization parameter, ${\rm log}\,\xi$, suddenly increases from 3.3 for observations with type-C QPOs to 3.7 in observations with type-B QPOs. Considering that $\xi=L/nR^{2}$, where $L$ is the ionizing luminosity, $n$ is the density of the disc and $R$ is the distance to the ionizing source, a sudden change of this parameter implies a change in the intrinsic properties of the source during the QPO transition period. As it is apparent from the Figure, the changes in all the parameters are very significant.

We also plot in Figure~\ref{fig:Pars_evo}, the evolution of the sizes, $L_{\rm 1}$ (small corona) and $L_{\rm 2}$ (large corona), and intrinsic feedback fraction, $\eta_{\rm{int,1}}$ and $\eta_{\rm{int,2}}$, of the two coronae and the temperature of the seed photon source that illuminates each corona, $kT_{\rm s,1}$ (=$kT_{\rm in}$) and $kT_{\rm s,2}$. In observations with type-C QPOs, the size of both coronae tends to increase, $L_{1}$ from $\sim$20--60\,km to $\sim$100\,km and $L_{2}$ from $\sim10^{4}$\,km to $\sim10^{5}$\,km. At the same time, the intrinsic feedback fraction of the large corona, $\eta_{\rm int,2}\sim$20--30\%, is always larger than that of the small corona, $\eta_{\rm int,1}\sim$5--15\%. Finally, the seed photons temperature of both coronae increases gradually, with $kT_{\rm s,2}\sim$0.3$-$0.5\,keV always being lower than $kT_{\rm s,1}\sim$0.4$-$0.6\,keV.  

Compared to the observations with type-C QPOs, in the observations with type-B QPOs the size of both coronae drops abruptly and significantly, $L_{1}$ form $\sim$100\,km to $\sim$20\,km and $L_{2}$ from $\sim10^{5}$\,km to $\sim10^{4}$\,km. During the transition from type-C to type-B QPOs, the intrinsic feedback fraction of the small corona increases from 15\% to 30\%, while that of the large corona decreases abruptly from 20--30\% to 0. During the transition the temperature of the seed photon source of the large corona is consistent with being constant, $kT_{\rm s,2}\sim$0.3\,keV, whereas the temperature of the seed photon source of the small corona, $kT_{\rm s,1}=kT_{\rm in}$, increases significantly from 0.65\,keV to 0.70\,keV. Here we do not give the values of the external heating rate $\delta \Dot{H}_{\rm ext}$ as they are not very well constrained (see Tables~\ref{tab:Fit_pars} and ~\ref{tab:Fit_pars_all}). For completeness, in Figure~\ref{fig:Pars_nu} we plot the parameters of the joint fits vs. QPO frequency. 

We note that while some parameters change abruptly at the time at which the type-C disappears and the type-B QPO appears (e.g., $kT_{\rm in}$, $\log{\xi}$, $L_2$ and $\eta_{\rm int,2}$), others evolve more or less continuously across the transition from the HIMS to the SIMS (e.g., $N_{\rm disk}$ and $\Gamma$). In fact, some of the parameters (e.g., $\log{\xi}$ and $L_2$) appear to be changing already in the HIMS, just before the transition. This suggests that the state transition, from the HIMS to the SIMS, is not a discontinuous process, but it starts before the QPO switches from type-C to type-B.

\section{Discussion}
\label{sec:discussion}

We analyze a \textit{NICER} observation of MAXI J1820+070 during the transition from the hard-intermediate (HIMS) to the soft-intermediate state (SIMS). A type-C quasi-periodic oscillation (QPO) with a frequency that increases from 4.4\,Hz up to 7.7\,Hz in the HIMS disappears at the time of the transition to the SIMS and is replaced by a type-B QPO at 4.2\,Hz that decreases to 3.1\,Hz as the source evolves further. We present the first comparison of the rms and lag spectra of the low-frequency QPO in a black hole X-ray binary during such state transition. The rms and lag spectra of the type-C QPO before, and the type-B QPO after, the transition are consistent with being the same above $\sim 1.5-2$\,keV but, remarkably, differ significantly below that energy. From fits to the energy spectra of the source and the rms and lag spectra of the QPOs in each of the 13 orbits in this observation with the time-dependent Comptonization model {\tt vkompth} \citep[][]{Karpouzas2020, Bellavita2022}, we infer for the first time the physical and geometrical properties of the corona in a black hole candidate (BHC) during the transition from the type-C to type-B QPOs. 

The data require two physically coupled coronae, a large one, with a size of $10^4-10^5$ km, that dominates the rms and lag spectra of the QPO and a small one, with a size of $\sim 20-100$ km, that impacts on the time-averaged spectra of the source. We find that in the HIMS $\sim$5--15\% of the flux of the small corona and $\sim$20--30\% of the flux of the large corona return to the disc. In the HIMS, while the returning flux of the small corona increases to 20--30\%, that of the large corona notably drops to zero in the SIMS.

These results indicate that a larger part of the accretion disc is enshrouded by the small corona in the SIMS than in the HIMS even though the corona is smaller, suggesting that the inner disc radius extends closer to the BH in the SIMS than in the HIMS. The high fraction of returning flux of the large corona in the HIMS suggests that in this state this corona covers a big portion of the accretion disc. On
the contrary, the fact that the large corona is as large in the SIMS as it was in the HIMS, but the returning flux is zero, indicates that in this state this corona does not cover the disc and therefore must extend vertically. This change in the geometry of the large corona coincides with the sudden increase of the radio emission of the source \citep[][]{Bright2020}, providing evidence of a connection between the X-ray corona and the radio jet in MAXI J1820+070, similar to what was observed in GRS 1915+105 \citep[][]{Mendez2022} and MAXI J1535--571 \citep[][]{Zhang2022}.

In the next subsections we first discuss the rms and lag spectra of the type-C and type-B QPOs, and the evolution of the corona properties. Subsequently, we examine our results in the context of models of the low-frequency QPO, and discuss the corona geometry during the state transition.

\begin{figure*}
\centering
\includegraphics[width=\textwidth]{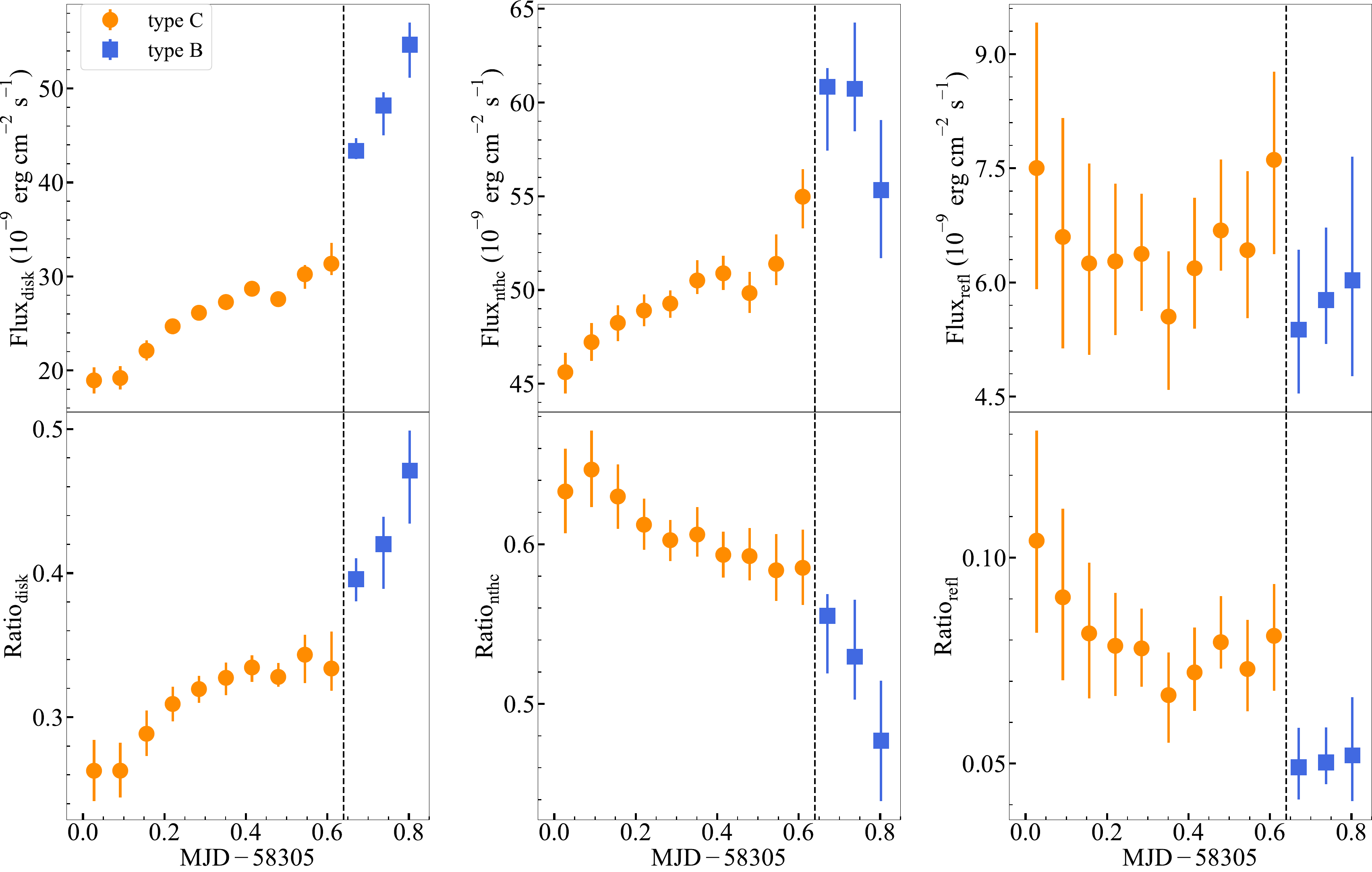} \\ 
\caption{Top panel: From left to right, unabsorbed flux of the, {\tt diskbb}, {\tt nthComp} and {\tt relxillCp} components in MAXI J1820+070 in the 0.5--10\,keV band. Bottom panel: The ratio of the flux of each component to the total flux.
\label{fig:flux}}
\end{figure*}

\subsection{The rms and lag spectra of the type-C and type-B QPOs}
Both for the type-C and type-B QPOs, the rms amplitude increases with energy up to the highest energy, similar to what is observed in other BHs, e.g., GX 339--4 \citep{Zhang2017}, GRS 1915+105 \citep{Zhang2020, Karpouzas2021}, MAXI J1535+571 \citep{Huang2018, Zhang2022, Zhang2023}, and MAXI J1348--630 \citep{Belloni2020,Alabarta2022,LiuHX2022}. The large rms amplitude at energies above 10\,keV \citep[in some cases the rms amplitude of the QPO is $\sim$12\% at 100--200\,keV;][]{Ma2021} implies that the variability must be produced by the corona, given that neither the disc nor the reflection component contributessignificantly to the emission at those energies \citep[e.g.,][]{Gilfanov2010,LiuHX2022}. 

We find that at energies above $\sim$2\,keV the rms spectrum of the two types of QPOs is similar, or the rms amplitude of the type-B QPO is somewhat higher than that of the type-C QPO. At energies below $\sim 1.5-2$\,keV, however, the rms amplitude of the type-C QPO is larger than 1\%, whereas that of the type-B QPOs is as low as 0.5\%, and significantly lower than that of the type-C QPO (see the left panel of  Figure~\ref{fig:RMS-LAG_pha}). In the time-dependent Comptonization model this is expected if one considers that $\Gamma$ is smaller and the feedback fraction of the large corona, $\eta_{2}$, is larger in the HIMS than in the SIMS (see Tables~\ref{tab:Fit_pars} and \ref{tab:Fit_pars_all}) because, if the rms amplitude and lags of the QPO are due to the variability of the Comptonized component, these trends lead to a lower variability at low energies \citep[see Figures 2 and 3 in][]{Bellavita2022}. At high energies, on the other hand, variable Comptonization predicts that the rms amplitude increases as $\Gamma$ increases \citep[see again][]{Bellavita2022}, consistent with the (slightly) higher rms amplitude of the type-B than of the type-C QPO above $\sim$2\,keV. 

As in other BHCs, the phase-lag spectra of both type-C and type-B QPOs in MAXI J1820+070 are ``U'' shaped. E.g., \citet[][]{Alabarta2022} reported that the lags of the type-C QPO in MAXI J1348--630 decreased abruptly with energy in the range 0.5--1.0\,keV and increased again as the energy increased further above that. A similar shape was reported for the phase-lag spectra of the type-C QPOs in MAXI J1535--571 \citep{Zhang2022} and the type-B QPOs in MAXI J1348--630 \citep[][]{Belloni2020,Garcia2021}, and GX 339--4 \citep{Peirano2022b}. 

Comparing the lag spectra of the type-C and type-B QPOs in MAXI J1820+070, we find a significant difference in the phase lags at energies below $\sim 1.5-2$\,keV, but almost no differences above that: The lag of the type-B QPO at the lowest energy band, 0.5\,keV, with respect to the minimum lag at 2\,keV is $\sim$1.4\,rad, whereas for the type-C QPO the lag is significantly lower, $\sim$0.2\,rad (see the right panel of Figure~\ref{fig:RMS-LAG_pha}). In the context of variable Comptonization, the lower the feedback fraction the larger the magnitude of the lags at lower energies \citep[other parameters have a smaller impact on the lags, see Figure 2 and 3 of ][]{Bellavita2022}. The difference of the lag spectra of the type-C and type-B QPOs at energies below 2\,keV are therefore consistent with the drop of the feedback fraction of the large corona in the fits, from $\eta_{\rm int,2}\sim 0.2-0.3$ in the HIMS (type-C QPO) to $\eta_{\rm int,2} =0$ in the SIMS (type-B QPO). Moreover, \citet{DeMarco2021} reported that during the state transition the thermal reverberation lags amplitude suddenly increased as the frequency decreased; they proposed that these changes were associated with the appearance of relativistic jet ejections. 

In addition, there is a marginally significant feature around 5--7\,keV in the phase-lag spectra of some type-C QPOs but not in the phase-lag spectra of the type-B QPOs (see the right panel of Figure~\ref{fig:RMS-LAG_pha}), which may correspond to the Fe K$\alpha$ line. \citet{Zhang2017} reported a similar feature in the phase-lag spectra of the type-C QPO of GX 339--4, and speculated that it could be associated with the reflection component. We note that, given the quality of the current data, there are no significant residuals in the lags in the range 5--7\,keV when we fit them with the time-resolved Comptonization model. 

\subsection{Coronae properties}
\label{subsec:coronae}

From the values of $\Gamma$ and $kT_{\rm e}$ we can compute the optical depth of the two coronae \citep{Zdziarski1996}:
\begin{equation}
\tau=\sqrt{\frac{9}{4}+\frac{3 m_{\rm e}c^{2}}{kT_{\rm e}\left(\Gamma-1\right)\left(\Gamma+2\right)}}-\frac{3}{2}.
\end{equation}
(Since we linked $\Gamma_1=\Gamma_2$ and $kT_{\rm e,1}=kT_{\rm e,2}$ during the fits, both coronae will have the same optical depth.). In the transition from type-C to type-B QPOs, $\Gamma$ increases from 2.4 to 2.9, which corresponds to a decrease of $\tau$ from 1.4 to 1.0. Since $kT_{\rm e}$ is fixed at 40\,keV in our fits because there is no cutoff in the spectrum within \textit{NICER}'s passband, the values of $\tau$ are likely upper limits.

As the source transition from HIMS to SIMS the Compton parameter, $y = (4kT_{\rm e}/m_{\rm e}c^{2}) \max{(\tau,\tau^{2})}$ \citep[][]{Zeldovich1969,Shapiro1976} decreases from $\sim$0.6 to $\sim$0.3. These values indicate that the Comptonization spectrum is in the unsaturated regime, which is consistent with the assumptions of the time-dependent Comptonization model.

The total flux of the source as well as the fluxes of the disc and Comptonization components, increase during the HIMS to SIMS, whereas the flux of the reflection component remains constant in the HIMS and decreases slightly during the transition (see Figure~\ref{fig:flux}). Furthermore, the ionization parameter increases accompanied by a softening of the time-averaged spectrum \citep[see also, ][]{Basak2017,Steiner2017}. These changes may be associated with changes in the geometry of the corona. 

Figure~\ref{fig:Pars_evo} (see also \ref{fig:Pars_nu}) shows the evolution of all parameters as the source transitions from the HIMS to the SIMS. For the small corona both the size and the feedback fraction are more or less constant in the HIMS; in the SIMS the feedback fraction appears to increase, while we only find upper limits to the size of the corona. For the large corona, in the HIMS the size and intrinsic feedback fraction increase with time, whereas in the SIMS, while the size is consistent with being constant at similar values as those in the HIMS at the start of the observation, the intrinsic feedback fraction is zero.

Any description of the evolution of the corona in this observation of MAXI J1820+070 needs to explain the fact that, while the size of the large corona is more or less the same at the start of the HIMS and in the SIMS, the feedback fraction is very different in the two states.

It is easy to imagine that a large fraction of the corona photons will impinge onto the disc, as it happens in the HIMS where the feedback is $\sim 20$\%, if the disc is enshrouded by the corona. On the contrary, it is very difficult to explain how in the SIMS a more or less equally large corona as in the HIMS would extend over the disc surface and yet give zero feedback. Considering the relatively small intrinsic feedback of such a large corona, especially in the SIMS, it is likely that the assumption of spherical symmetry of the corona in the {\tt vkompth} model does not apply, and the best-fitting values should be taken as characteristic corona sizes rather than as the actual radius of a spherical corona. The size of the large corona and very low feedback suggest that in the SIMS this corona extends vertically.

Similar situations have also been reported in other sources. For instance, for the type-C QPO in GRS 1915+105, \citet{Karpouzas2021}, \citet[][]{Mendez2022} and \citet[][]{Garcia2022} found that the size of the corona decreased from $\sim10^{3}$\,km to $\sim10^{2}$\,km with a large feedback fraction when the QPO frequency decreased from $\sim$6\,Hz to $\sim$2\,Hz, while the corona size increased again to $\sim2\times10^{4}$\,km, but this time with zero feedback, when the QPO frequency decreased further from $\sim$2\,Hz to $\sim$0.5\,Hz. \citet[][]{Mendez2022} interpreted these results as a change of the geometry of the corona, from a torus-like corona covering a large fraction of the accretion disc when the QPO frequency was larger than $\sim$2\,Hz, to a vertically extended, jet-like, corona when the QPO frequency was below $\sim$2\,Hz. Similar interpretations of the geometry of the corona based on the type-B QPO in MAXI J1348--630 and the type-C QPOs in MAXI J1535+571 were given by \citet{Garcia2021} and \citet{Zhang2022}, respectively.

\subsection{Other models of the low frequency QPOs}

\subsubsection{Models of the type-C QPO}
\citet{Ingram2009} proposed that the type-C QPOs originate from the Lense–Thirring (LT) precession of a hot inner flow, located inside the truncated accretion disc, that periodically illuminates the (steady) accretion disc during the QPO phase. \citet[][]{Ingram2016} reported that  $\Gamma$  and the centroid energy of the iron line were modulated during the QPO cycle ($\succsim3\sigma$ confidence) in the BHB H1743--322, while \citet[][]{Nathan2022} reported that $\Gamma$ and the reflection fraction were modulated during the QPO cycle ($\succsim3\sigma$ confidence) in the BHB GRS 1915+105. We note that the modulation of $\Gamma$ is also expected in the case of time-dependent Comptonization as a result of oscillations of the disc and corona temperatures even if, as in the case of the \texttt{vkompth} model \citep[][]{Karpouzas2020, Bellavita2022}, no explicit oscillations of $\Gamma$ are included. While LT precession can in principle account for the dependence of the rms amplitude \citep[][]{Motta2015} and the phase/time lags \citep[][]{vdEijnden2017} of the QPO upon the source inclination, recent results appear to challenge the LT idea at least in its current form. Specifically, \citet{Nathan2022} applied this model to the type-C QPO in GRS 1915+105 observed with \textit{NuSTAR} and \textit{XMM-Netwon}, and found that the model yields an uncomfortably small inner disc radius, $R_{\rm in}\sim$1.4\,$R_{\rm g}$, and a problematically long thermalization time scale, $t_{\rm th}\sim$70\,ms. They argue that such a long thermalization time scale could be an artifact of some simplifications of the model, e.g., a possible radial dependence of the thermalization time scale or the ionization parameter. According to \cite{Nathan2022}, other effects that could lead to these results are the possible shadowing of the illumination profile, or that the model ignores the light crossing delay in the disc illumination profile, which could affect the disc temperature profile. Recently, \citet{Salvesen2022} modeled thermalization as a random walk of photons in an $\alpha$-disc, and found that for most of the disc $t_{\rm th}<$1\,ms, much smaller than the values reported by \citet{Nathan2022}. 

Mass accretion rate fluctuations excited in each ring of the accretion disc and propagating toward the BH have been proposed as an explanation of the lags of the broad-band noise component in the PDS of these systems \citep[][]{Lyubarskii1997, Kotov2001,Arevalo2006}. This mechanism also explains the rms-flux relation in BHBs \citep[e.g.,][]{Uttley2001, Uttley2005}. \citet[][]{Ingram2013} combined the propagating mass accretion rate fluctuations and the LT precession of the inner flow and applied this idea to reproduce the power and lag spectra of the BHB XTE J1550--564. A recent study of MAXI J1820+070 shows that propagation of mass accretion rate fluctuation can explain the broad-band noise and the hard lags observed in this source but not the soft lags \cite{Kawamura2022}. On the other hand, \citet{Rapisarda2017} showed that this model fails to reproduce the power and cross spectra of the BHC XTE J1550--564, and concluded that these discrepancies are generic to the propagating fluctuations paradigm. 

A model with two accretion flows, a standard accretion disk (SAD) in the outer parts and a jet-emitting disk (JED) in the inner parts (SAD-JED) was proposed by \citet[][]{Ferreira1997}, and further developed by \citet[][]{Marcel2018}. In this model, the type-C QPOs are related to the transition radius between the SAD and the JED \citep[][]{Marcel2020}. In the SAD-JED model, the variability is due to the propagation of mass accretion rate fluctuation and therefore has the same issues explaining the soft lags.

Recently, \citet{Mastichiadis2022} proposed that the type-C QPOs in the BHXRB could arise from the coupling of the Comptonizing corona and the accretion disc. They suggested that, under certain circumstances, this coupled system would oscillate at a characteristic frequency that depends upon the fraction of the corona photons that return to the disc, the mass accretion rate, the size of the corona, and the black-hole mass. They found that for typical values of these parameters this characteristic frequency matches the range of frequencies of the type-C QPO in these systems. The essence of the proposal of \citet[][]{Mastichiadis2022} is the same as that of {\tt vkompth}, the time-dependent Comptonization model of \citet[][]{Karpouzas2020}; while \citet[][]{Mastichiadis2022} consider the disc-corona coupling to explain the frequency of the QPO, the {\tt vkompth} model uses the same coupling to explain the rms and phase-lag spectra of the QPO.
The QPO frequency predicted by \citet[][]{Mastichiadis2022} is
\begin{equation}
f_{0} \simeq 5\,{\rm Hz} \left(\frac{\eta_{\rm int}}{0.2}\right)^{1/2} \left(\frac{\dot{m}}{10^{-3}}\right)^{1/2} \left(\frac{L}{30 R_{\rm g}}\right)^{-3/2} \left(\frac{\xi f}{3}\right)^{-1/2} \left(\frac{M}{10 M_{\odot}}\right)^{-1},
\end{equation}
where $f_{0}$ is the QPO frequency, $\dot{m}$ is the accretion rate in Eddington units, $\eta_{\rm int}$ and $L$ are, respectively, the intrinsic feedback fraction and the corona size (in units of $R_{\rm g} = GM/c^2$) in the {\tt vkompth} model, $\xi$ is a factor that can be either $1/2$ or $3$ for, respectively, a maximally and a non-rotating black hole, $f$ is the fraction of the gravitational energy that is injected into the corona, and $M$ is the black-hole mass in solar masses. If we take $\eta_{\rm int} \sim 0.2$ and $L\sim800\,R_{\rm g}$ from our results, $M\sim8\,M_{\odot}$ \citep{Torres2020}, $\xi=1/2$ for a maximally spinning black hole \citep[e.g.,][]{Wang2021} and $f =0.1-0.2$, the QPO frequency is $f_{0} \simeq 2-8$\,Hz for $\dot{m}\sim 0.1-0.5\,\dot{M}_{\rm Edd}$ \citep[e.g.][but notice that due to the uncertainties in $\eta_{\rm int}$ and $L$ these frequencies have large errors]{Dunn2010}.

\subsubsection{Models of the type-B QPO}

\citet[][]{Stevens2016} analyzed the type-B QPO in the BHB GX 339--4 using phase-resolved spectroscopy, and they found that $\Gamma$ varies periodically during the QPO cycle. They interpreted this as the precession of the jet in this source. Subsequently, using Monte Carlo simulations \citet[][]{Kylafis2020} explained these variations of $\Gamma$ as the precession of the jet in GX 339--4, further supporting this idea. \citet{LiuHX2022} studied the type-B QPOs in MAXI J1348--630 and suggested that it also originates from the jet precession; they also attributed the disappearance of the type-B QPOs to instabilities of the disc-jet structure. 

If the type-B QPO is indeed produced by the precession of the jet, it is reasonable to expect a correlation between the properties (e.g., rms and lag) of type-B QPOs and the inclination of the jet with respect to the line of sight. The lag spectra in the 1--10\,keV band of MAXI J1348--630 \citep[][]{Garcia2021}, GX 339--4 \citep{Peirano2022b} and MAXI J1820+070 in this work show that, for the type-B QPOs the $\sim$2--3\,keV photons always lead the photons in other energy bands, and that the magnitude of lags is always larger below than above that energy. Given the similarities of the lag spectra of the type-B QPO in these three sources, the fact that the inclination of the jet in these sources are quite different ($\sim33^{\circ}$ for MAXI J1348--630, \citealt[][]{Mall2022}; $\sim35-45^{\circ}$ for GX 339--4, \citealt[][]{LiuHH2022b}; $\sim63^{\circ}$ for MAXI J1820+070, \citealt[][]{Torres2020}) appears to question this scenario.\\

\subsection{Evolution of the corona}
\label{subsec:model}

Using the combined results of the fits to the time-averaged spectra of the source and the rms and lag spectra of the QPO plus the quasi-simultaneous radio observation \citep[][see Figure~\ref{fig:radio_size2}]{Bright2020}, we propose a possible scenario during the transition from the HIMS to the SIMS in Figure~\ref{fig:model}. 

\subsubsection{State transition, radio flare and jet ejections}
\label{sec:QPO_trans}

We first checked and confirmed that the PDS of orbit 09 and orbit 10 with \textit{NICER} change significantly, indicating that the QPO transition occurs in this period. We find that the PDS of the whole orbit 09 shows a strong broad-band noise component, whereas the PDS of the whole orbit 10 shows a weak broad-band noise component (see Figure~\ref{fig:PDS_0910}). This indicates that in orbit 09, MAXI J1820+070 is still in the HIMS, whereas in orbit 10, it is already in the SIMS. The transition time must have therefore occurred between these two orbits, i.e., MJD $58305.66\pm{0.02}$.

Subsequently, we re-examined the results of the radio data during this state transition. \citet[][]{Bright2020} reported that, in addition to a core quenching with gradually decreasing flux, the observations of MAXI J1820+070 with AMI--LA showed a radio flare that peaked $\sim 0.3$ days after the time of the transition of type-C to type-B QPOs (see Figure~\ref{fig:radio_size2}), which they associated with the ejection of ballistic jets at around the time of the transition. Using the dynamic phase center tracking technique to fit all the VLBI data, \citet{Wood2021} found that the launch date was MJD $58305.60\pm{0.04}$ \citep[see][who found a similar launch date modelling a subset of the same data]{Bright2018,Espinasse2020}. From this it appears that the ejection that caused the flare on $\sim$ MJD 59306 may have been launched already during orbit 09, when the source was still in the HIMS with a type-C QPO, just before the transition to the SIMS with type-B QPO. It is worth noting that $L_2$, the size of the large corona, was already increasing before that date, and reached its maximum value during orbit 09 (see Fig.~\ref{fig:Pars_evo}).

\subsubsection{Corona geometry in the HIMS}
\begin{figure}
\centering
\includegraphics[width=\columnwidth]{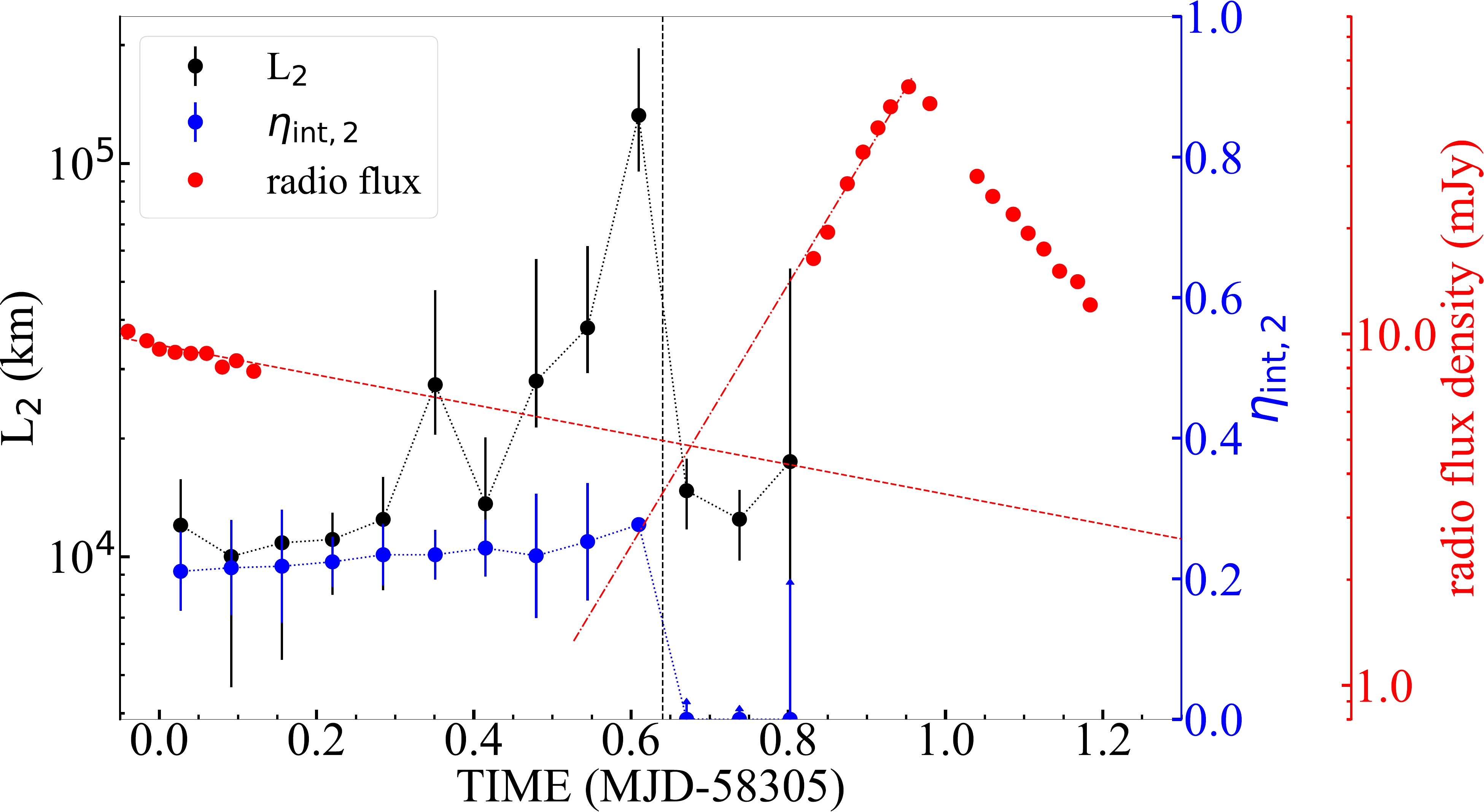} \\ 
\caption{The evolution of the size of the large corona and the radio flux of MAXI J1820+070 during the transition from type-C to type-B QPOs. The black, blue, and red points represent the size of the large corona, $L_{2}$, the intrinsic feedback fraction of the large corona, $\eta_{\rm{int,2}}$, and the AMI-LA radio flux, respectively. The black vertical line represents the QPOs transition time, and the red dashed and dashed-dotted lines represent the best-fitting evolution of, respectively, the radio emission of the core that decreases steadily during the period with type-C QPOs and the radio flare during the period with type-B QPOs. The radio data and lines showing the evolution of the radio emission are from \citet[][]{Bright2020}.
\label{fig:radio_size2}}
\end{figure}

\begin{figure}
\centering
\includegraphics[width=\columnwidth]{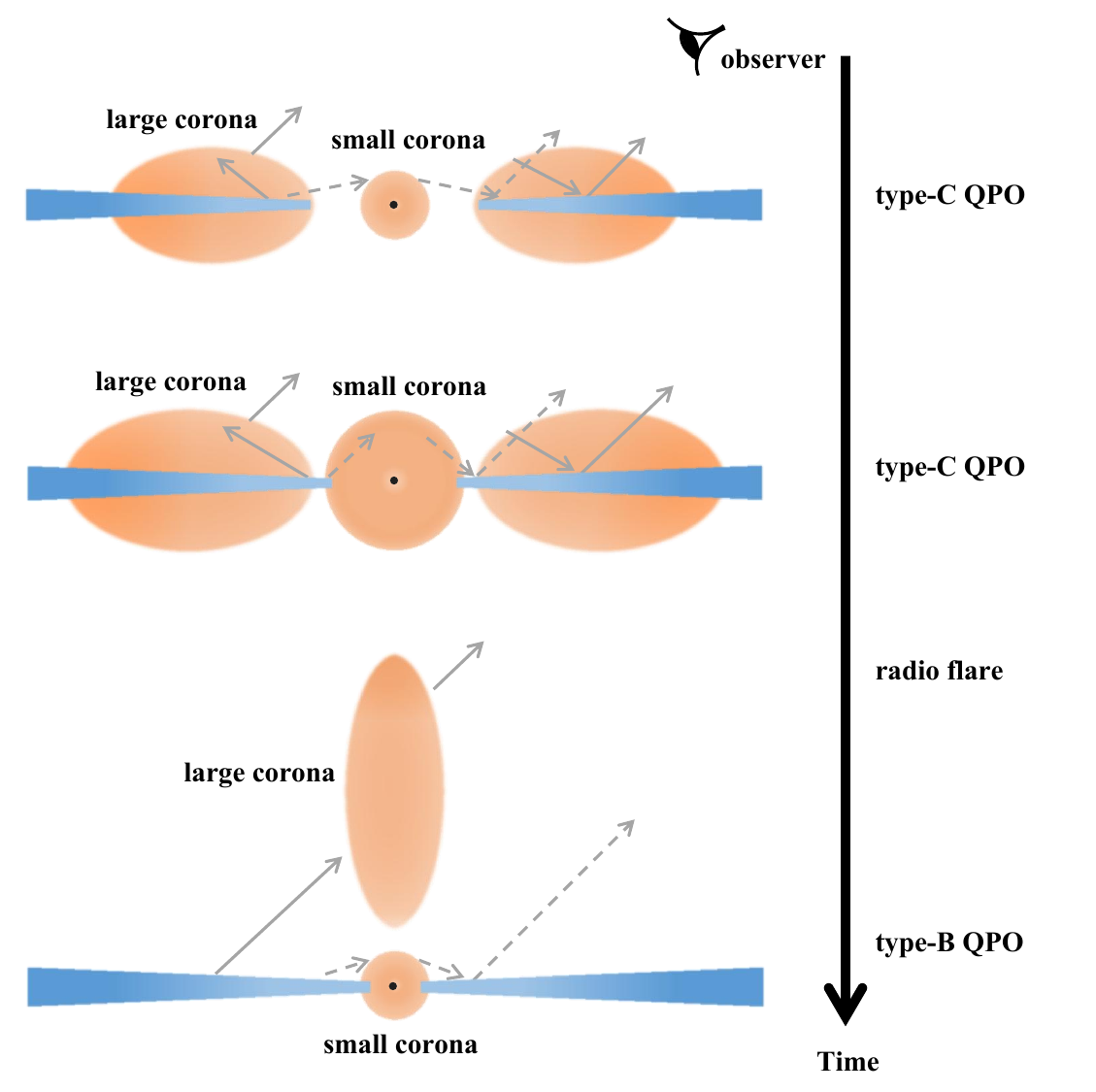} \\ 
\caption{Schematic picture of the corona evolution during the transition from type-C QPOs to type-B QPOs in MAXI J1820+070. In the type-C QPOs case, the small corona is located inside the inner disc radius and the seed photons come from the inner disc. The large corona extends horizontally covering the inner parts of the disc, and the seed photons for this corona come from the outer parts of the disc. The values of $\eta_{\rm{int,1}}$ and $\eta_{\rm{int,2}}$ indicate that both coronae illuminate the disc (top panel). As the outburst evolves, the size of both coronae increases while the inner disc radius moves inwards (middle panel). The radio emission increases rapidly at the transition from the type-C to the type-B QPOs \citep[see Figure~\ref{fig:radio_size2} and][]{Bright2020}. In the type-B QPOs case the small corona is still located inside the inner disc radius, while the large corona extends vertically, consistent with the detection of the relativistic ejecta (bottom panel). 
\label{fig:model}}
\end{figure}

In the HIMS, the centroid frequency of the QPO increases from 4.4\,Hz to 7.7\,Hz (see Section 4.2 and Figures~\ref{fig:Pars_evo} and \ref{fig:radio_size2} for the evolution of the parameters of the coronae). The temperature of the seed photons that illuminate the large corona shows an increasing trend while being always lower than the temperature of the seed photon source that illuminates the small corona. These results suggest that the small corona is located inside the inner edge of the accretion disc and is illuminated by the central parts of the disc, while the large corona covers a large fraction of the accretion disc and is illuminated by a cooler part of the disc (top and middle panels of Figure~\ref{fig:model}). 

The Imaging X-ray Polarimetry Explorer \citep[IXPE;][]{Weisskopf2022} observed that Cyg X--1 in the non-thermal component dominated state has a polarization degree $\sim$4\% in the 2--8\,keV energy band at a position angle of $\sim-21^{\circ}$ \citep[][]{Krawczynski2022}, consistent with the position angle of the jet on the sky \citep{Miller-Jones2021}. Modeling of these results indicates that in Cyg X--1 in the non-thermal dominated state, Comptonization takes place in a sandwich corona located above and below the disc and extending laterally. The proposed geometry of MAXI J1820+070 in the HIMS that we deduce from our X-ray analysis is consistent with that of Cyg X--1 deduced from those X-ray polarimetric measurements.

Steady, optically thick, radio emission is observed in this source in the HIMS \citep{Bright2020}, and other sources both in the LHS and the HIMS \citep[e.g.,][]{Corbel2003,Fender2004}. That the existence of this emission does not correspond to a vertically extended corona \citep[see our results here and][]{Krawczynski2022}, suggests that the radio emission is not from material ejected into a narrow jet, but comes from a mildly relativistic outflow that originates from a relatively large radius in the disc \cite[e.g.][]{Kylafis2018,Marcel2018,Reig2021}.

\subsubsection{Corona geometry in the SIMS}

In the SIMS, the centroid frequency of the QPO decreases from 4.2\,Hz to 3.1\,Hz. The parameters of two coronae during SIMS, and the comparison with the HIMS, are shown in Section 4.2 and Figures~\ref{fig:Pars_evo} and \ref{fig:radio_size2}. The parameters of the coronae indicate that, as in the HIMS, in the SIMS the small corona is also located inside the inner edge of the disc, whereas the vanishing small intrinsic feedback fraction indicates that the geometry of the large corona changes significantly. From all this we interpret that the radius of the large corona is in fact a characteristic size and that the large corona is vertically extended \citep[see also,][]{Mendez2022, Zhang2023}. As in the HIMS, in the SIMS the temperature of the seed-photon source of the small corona is higher than that of the large corona. Moreover, the temperature of the seed-photon source of the small corona increases further in the SIMS compared to the HIMS, whereas that of the large corona decreases from the HIMS to the SIMS. This indicates that in the SIMS the large corona is illuminated from the parts farther away from the BH than in the HIMS, consistent with the proposed geometry.

We compare our results to the radio observation during the QPO transition period. As shown in Figure~\ref{fig:radio_size2} and described in Section~\ref{sec:QPO_trans}, the launch date of the discrete ejection that produced the radio flare, MJD $58305.60\pm{0.04}$, is consistent with the QPO transition time, MJD $58305.66\pm{0.02}$. Both \citet{Espinasse2020} and \citet{Wood2021} suggest that the radio flare is associated with the occurrence of type-B QPOs. \citet{Homan2020} also reported the relation between the type-B QPOs and the radio flare, that the appearance of type-B QPOs was accompanied by an X-ray flare in the 7--12\,keV energy band, and that a strong radio flare appeared 2--2.5 hr after the type-B QPOs was observed. They interpreted the delay of the radio flare with respect to the X-ray flare and the QPO transition as the time for the ejecta to catch up with the material of the previous slower ejecta, or the time for the ejecta to interact with the interstellar medium. Based on our and the above radio results, we suggest that the large corona is a jet-like corona extending vertically with no photons feeding back to the disc, and the small corona is located inside the inner disk radius (bottom panel of Figure~\ref{fig:model}).

\subsubsection{Comparison with other corona geometries in MAXI J1820+070}

From an X-ray spectroscopic analysis of \textit{NICER} data, \citet[][]{Zdziarski2021} also proposed that, in the hard state, the corona of MAXI J1820+070 consists of two Comptonization regions. These authors find that a hotter corona, which is located farther away from the accretion disc, dominates the narrow reflection component and is responsible for the reverberation lags, whereas a cooler component located closer to the BH is responsible for the broader reflection component. Subsequent analysis of the energy spectra of the Lorentzian functions used to fit the PDS of the source also suggested the existence of two Comptonized regions with different temperatures and optical depths \citep{Dzieak2021}. 

Broadband variability studies also indicate that the accretion flow structure changed during the state transition in MAXI J1820+070. Fits of the lag-energy spectra in a broad frequency range using the reverberation model {\tt reltrans} \citep{Ingram2019b} showed that the corona expanded vertically and the height\footnote{Because the {\tt reltrans} model assumes that the corona is a point source, the model gives the height of this point source above the accretion disc instead of the size of the corona.} of the corona increased during the state transition of MAXI J1820+070 \citep{Wang2021}, reaching corona heights comparable to the size of the large corona that we find here \citep[see also][]{Lucchini2023}. Through timing-spectral analysis, \citet{DeMarco2021} also reported a dramatic change in the structure of the innermost accretion flow during the state transition, although they suggested that the transition of the source was accompanied by a decrease of the disc inner radius rather than an increase of the corona height.

\section{Conclusions}
\label{sec:conclusion}

We analyze \textit{NICER} data of 13 orbits during the state transition of MAXI J1820+070 from the HIMS to the SIMS. We use the time-dependent Comptonization model, {\tt vkompth}, to fit jointly the time-averaged spectra of the source and the rms and lag spectra of the QPOs to study the spectral-timing evolution of this source. Our main findings are:

\begin{itemize}
\item[1.]
We detect a type-C QPO with a centroid frequency increasing from 4.4 to 7.7\,Hz in the HIMS that evolves to a type-B QPO with a centroid frequency that decreases from 4.2 to 3.1\,Hz in the SIMS.
\end{itemize}

\begin{itemize}
\item[2.]
The rms and lag spectra of both the type-C and type-B QPOs are consistent with being the same above $\sim 1.5-2$\,keV. At energies below $\sim 1.5-2$\,keV the type-C QPOs show a larger rms amplitude and a lower magnitude of the phase lag than the type-B QPOs.
\end{itemize}

\begin{itemize}
\item[3.]
The data can be fitted with a time-dependent Comptonization model of two physically coupled coronae: In the HIMS the system consists of a small corona of $\sim$20--100\,km with an intrinsic feedback fraction of $\sim$5--15\%, and a large corona of $\sim10^{4}-10^{5}$\,km with an intrinsic feedback fraction of $\sim$20--30\%. In the SIMS, the size of the small corona decreases to $\sim$20\,km with an intrinsic feedback fraction of $\sim$20--30\%, while the size of the large corona is $\sim10^{4}$\,km with an intrinsic feedback fraction of 0. At the same time, the inner radius of the disc inferred from the \textit{diskbb} component decreases during the transition.
\end{itemize}

\begin{itemize}
\item[4.]
From the fitting results from \textit{NICER} and the radio data previously published, we propose a possible evolution of the  corona geometry during the state transition. In the HIMS, the system consists of a large and horizontally extended corona covering the inner parts of the accretion disc, and a small corona located inside the inner edge of the disc. In the SIMS, the small corona is still located inside the inner edge of the disc, whereas the large corona is vertically extended and is associated with the material that produced the radio flare.
\end{itemize}

\section*{Acknowledgements}


We thank the anonymous referee for useful comments that helped us improve the paper. We thank Kevin Alabarta for providing a script that helped process the \textit{NICER} data. We thank Thomas Russell and Dave Russell for useful comments. This work is supported by the  China Scholarship Council (CSC 202104910402). M.M. acknowledges the research programme Athena with project number 184.034.002, which is (partly) financed by the Dutch Research Council (NWO). FG is a CONICET researcher. FG acknowledges support by PIP 0113 (CONICET), PICT-2017-2865 (ANPCyT) and PIBAA 1275 (CONICET). YZ acknowledges support from China Scholarship Council (CSC 201906100030). MM and FG thank the Team Meeting at the International Space Science Institute (Bern) for fruitful discussions. This research has made use of NASA’s Astrophysics Data System.

\section*{Data Availability}


The data used in this article are available in the HEASARC database (\url{https://heasarc.gsfc.nasa.gov}). The {\sc vkompth} model is publicly available in a Github repository (\url{https://github.com/candebellavita/vkompth}). The corner plots were created using {\sc tkXspecCorner} (\url{https://github.com/garciafederico/pyXspecCorner}).



\bibliographystyle{mnras}
\bibliography{ref} 




\appendix

\section{Fit with {\tt relxilLCp} v2.3}
\label{Appendix-A}
During our spectral fitting using {\tt relxillCp} v1.4.3, we find that the iron abundance is always $A_{\rm Fe}>9.91$. Similar values of the iron abundance have been reported in the past from fits with {\tt relxill} to the spectra of Cyg X--1 \citep{Tomsick2018} and GX 339--4 \citep{Jiang2019}, and were attributed to limitations of the model that at the time was computed for a low disc density $N = 10^{15}$ cm$^{-3}$. Since version 2.3 of {\tt relxill} includes tables with $N$ ranging from $10^{15}$ cm$^{-3}$ to $10^{20}$ cm$^{-3}$, we fit the data using the new version and find a best-fitting disc density $N \sim 10^{17}$~cm$^{-3}$ and $A_{\rm Fe} \simeq5$. Other parameters of the fit are consistent with the ones we find using version 1.4.3 within errors. Since in this paper we are not interested in the iron abundance, we give the fits with version 1.4.3 because the temperature of the seed-photon source is fixed at $kT_{\rm bb} = 0.05$ keV, whereas in version 2.3 it is fixed at $kT_{\rm bb} = 0.01$~keV.

\section{Table with all fitting parameters and extra plots}

\begin{figure*}
\centering
\includegraphics[width=\textwidth]{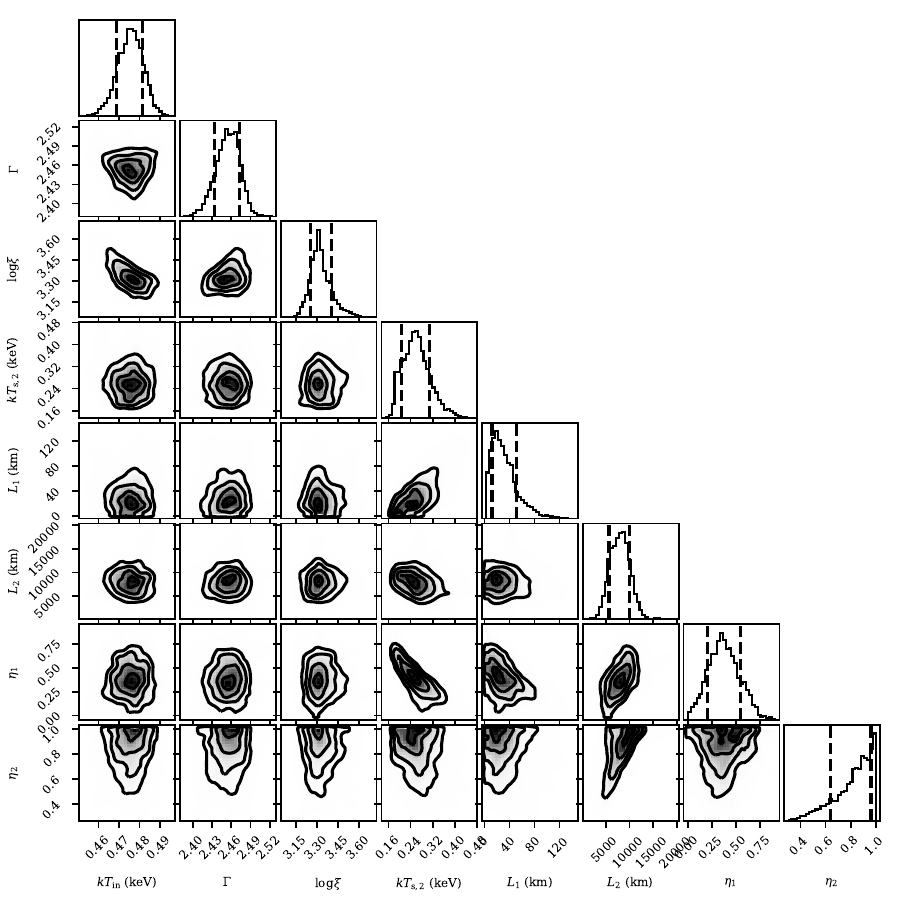} \\ 
\caption{Representative corner plot of the parameters of MAXI J1820+070 for an observation with the type-C QPO (orbit 01). The contours indicate the 1, 2 and 3-$\sigma$ levels for two parameters, while the vertical lines indicate the 1-$\sigma$ confidence range for one parameter.
\label{fig:Cont1}}
\end{figure*}

\begin{figure*}
\centering
\includegraphics[width=\textwidth]{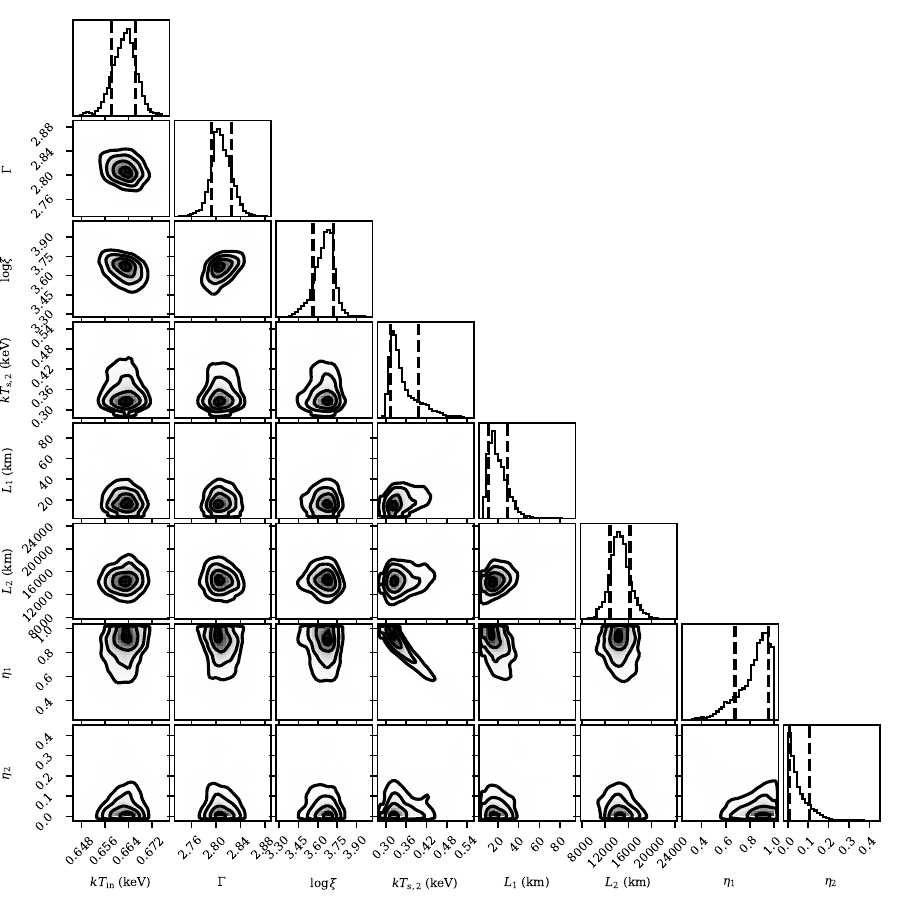} \\ 
\caption{Same a Figure~\ref{fig:Cont1} for an observation with the type-B QPO (orbit 10).
\label{fig:Cont10}}
\end{figure*}

\begin{figure*}
\centering
\includegraphics[width=\textwidth]{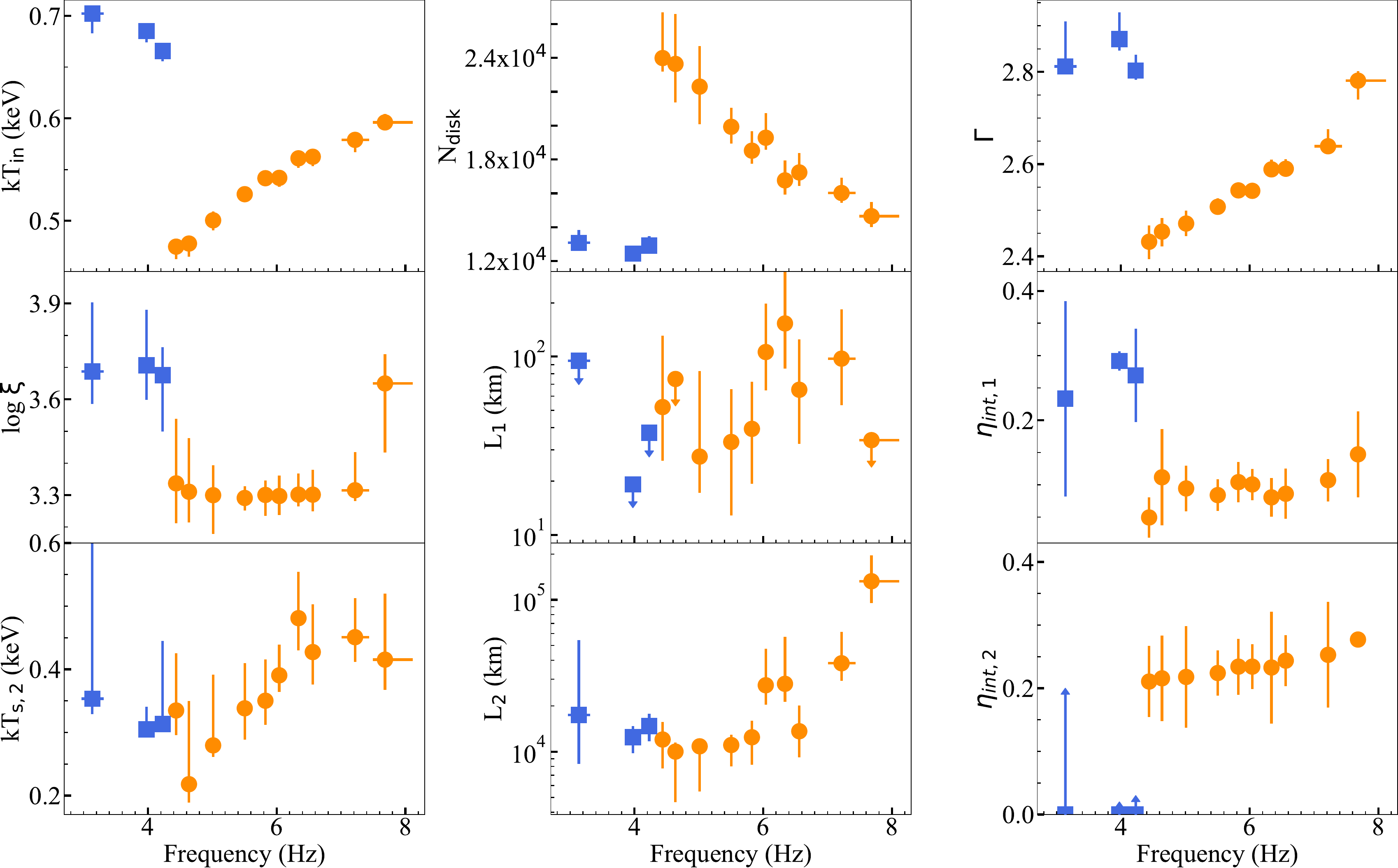} \\ 
\caption{Best-fitting parameters as a function of QPO frequency during the transition from type-C to type-B QPOs of MAXI J1820+070. The parameters and the colors are the same as in Figure~\ref{fig:Pars_evo}. 
\label{fig:Pars_nu}}
\end{figure*}



\begin{table*}
    \renewcommand\arraystretch{1.5}
	\centering
	\caption{Best-fitting parameters of the 13 orbits during the transition from type-C (orbit 00$-$obit 09) to type-B (orbit 10$-$orbit 12) QPO of MAXI J1820+070. Uncertainties are at 90\% confidence levels.}
	\label{tab:Fit_pars_all}
	\begin{threeparttable}
	\resizebox{\textwidth}{!}{
	\begin{tabular}{lcccccccccccccc} 
	\hline
	Component & Parameter & orbit 00 & orbit 01 & orbit 02 & orbit 03 & orbit 04 & orbit 05 & orbit 06 & orbit 07 & orbit 08 & orbit 09 & orbit 10 & orbit 11 & orbit 12 \\
	\hline
	
    TBfeo& $N_{\rm H}\,(10^{20}\,\rm{cm^{-2}})$ & [4.2]$^{a}$ & [4.2]$^{a}$ & [4.2]$^{a}$ & [4.2]$^{a}$ & [4.2]$^{a}$ & [4.2]$^{a}$ & [4.2]$^{a}$ & [4.2]$^{a}$ & [4.2]$^{a}$ & [4.2]$^{a}$ & [4.2]$^{a}$ & [4.2]$^{a}$ & [4.2]$^{a}$ \\ 
    diskbb & $kT_{\rm in}$ (keV) & $0.475_{-0.012}^{+0.004}$ & $0.478_{-0.013}^{+0.007}$ & $0.501\pm{0.010}$ & $0.526_{-0.006}^{+0.003}$ & $0.542\pm{0.005}$ & $0.542_{-0.009}^{+0.004}$ & $0.563_{-0.009}^{+0.005}$ & $0.561_{-0.009}^{+0.006}$ & $0.579_{-0.012}^{+0.005}$ & $0.596\pm{0.008}$ & $0.666_{-0.010}^{+0.003}$ & $0.685_{-0.011}^{+0.005}$ & $0.702_{-0.019}^{+0.001}$  \\ 
     & ${N_{\rm disk}}^{b}\,(10^{4})$ & $2.40_{-0.08}^{+0.27}$ & $2.4_{-0.2}^{+0.3}$ & $2.2\pm{0.2}$ & $1.99\pm{0.11}$ & $1.85_{-0.08}^{+0.12}$ & $1.93_{-0.07}^{+0.15}$ & $1.72_{-0.08}^{+0.12}$ & $1.68_{-0.08}^{+0.12}$ & $1.60_{-0.06}^{+0.09}$ & $1.47\pm{0.07}$ & $1.29_{-0.03}^{+0.06}$ & $1.24\pm{0.04}$ & $1.31_{-0.02}^{+0.08}$  \\ 
     & ${F_{\rm disk}}^{*}\,(\rm{10^{-9}~ergs~cm^{-2}~s^{-1}})$ & $18.9\pm{1.4}$ & $19.2\pm{1.3}$ & $22.1\pm{1.1}$ & $24.7\pm{0.8}$ & $26.1\pm{0.6}$ & $27.3\pm{0.8}$ & $28.7\pm{0.6}$ & $27.6\pm{0.5}$ & $30.2_{-1.6}^{+1.0}$ & $31.4_{-1.2}^{+2.2}$ & $43.4_{-0.9}^{+1.4}$ & $48.2_{-3.2}^{+1.4}$ & $55\pm{3}$  \\      
    nthComp & $\Gamma$ & $2.43\pm{0.04}$ & $2.45\pm{0.03}$ & $2.47\pm{0.03}$ & $2.51\pm{0.02}$ & $2.543\pm{0.016}$ & $2.54\pm{0.02}$ & $2.590_{-0.010}^{+0.020}$ & $2.589_{-0.012}^{+0.021}$ & $2.64\pm{0.03}$ & $2.78\pm{0.03}$ & $2.80\pm{0.03}$ & $2.87_{-0.03}^{+0.06}$ & $2.812_{-0.018}^{+0.097}$  \\ 
     & $kT_{\rm e}$ (keV) & [40]$^{a}$ & [40]$^{a}$ & [40]$^{a}$ & [40]$^{a}$ & [40]$^{a}$ & [40]$^{a}$ & [40]$^{a}$ & [40]$^{a}$ & [40]$^{a}$ & [40]$^{a}$ & [40]$^{a}$ & [40]$^{a}$ & [40]$^{a}$  \\ 
     & $N_{\rm nthComp}$ & $9.8\pm{0.3}$ & $10.2\pm{0.4}$ & $10.2\pm{0.3}$ & $10.19_{-0.12}^{+0.25}$ & $10.2\pm{0.3}$ & $10.5\pm{0.3}$ & $10.5\pm{0.3}$ & $10.3\pm{0.3}$ & $10.6_{-0.3}^{+0.6}$ & $11.7\pm{0.5}$ & $11.7_{-0.3}^{+0.8}$ & $11.9_{-0.4}^{+1.1}$ & $9.88_{-0.09}^{+1.72}$  \\
     & ${F_{\rm nthComp}}^{*}\,(\rm{10^{-9}~ergs~cm^{-2}~s^{-1}})$ & $45.6\pm{1.1}$ & $47.2\pm{1.0}$ & $48.2\pm{1.0}$ & $48.9\pm{0.9}$ & $49.3\pm{0.7}$ & $50.5_{-0.7}^{+1.1}$ & $50.9\pm{0.9}$ & $49.8\pm{1.1}$ & $51.4_{-1.1}^{+1.6}$ & $55.0\pm{1.6}$ & $60.8_{-3.4}^{+1.0}$ & $61\pm{3}$ & $5.5\pm{0.4}$  \\      
    relxillCp & $\alpha1$ & $2.09_{-0.13}^{+0.18}$ & $2.23_{-0.14}^{+0.20}$ & $2.13\pm{0.15}$ & $2.13\pm{0.10}$ & $2.12\pm{0.08}$ & $2.09_{-0.12}^{+0.09}$ & $2.13\pm{0.11}$ & $2.28\pm{0.13}$ & $2.19\pm{0.11}$ & $2.18\pm{0.12}$ & $2.06_{-0.16}^{+0.10}$ & $2.12\pm{0.13}$ & $2.17_{-0.31}^{+0.08}$  \\ 
    & $a_{\rm *}$ & [0.998]$^{a}$ & [0.998]$^{a}$ & [0.998]$^{a}$ & [0.998]$^{a}$ & [0.998]$^{a}$ & [0.998]$^{a}$ & [0.998]$^{a}$ & [0.998]$^{a}$ & [0.998]$^{a}$ & [0.998]$^{a}$ & [0.998]$^{a}$ & [0.998]$^{a}$ & [0.998]$^{a}$  \\  
    & $i$ ($^{\circ}$) & [67.3]$^{a}$ & [67.3]$^{a}$ & [67.3]$^{a}$ & [67.3]$^{a}$ & [67.3]$^{a}$ & [67.3]$^{a}$ & [67.3]$^{a}$ & [67.3]$^{a}$ & [67.3]$^{a}$ & [67.3]$^{a}$ & [67.3]$^{a}$ & [67.3]$^{a}$ & [67.3]$^{a}$ \\     
     & ${\rm log}\,\xi\,(\rm{log[erg\,cm\,s^{-1}}])$ & $3.34_{-0.13}^{+0.20}$ & $3.31_{-0.10}^{+0.17}$ & $3.30_{-0.12}^{+0.09}$ & $3.29\pm{0.04}$ & $3.30_{-0.07}^{+0.05}$ & $3.30\pm{0.06}$ & $3.30_{-0.05}^{+0.08}$ & $3.30_{-0.04}^{+0.07}$ & $3.32_{-0.03}^{+0.12}$ & $3.65_{-0.22}^{+0.09}$ & $3.68_{-0.18}^{+0.09}$ & $3.71_{-0.11}^{+0.17}$ & $3.69_{-0.10}^{+0.22}$  \\ 
     & $A_{\rm Fe}$ (solor) & [10]$^{a}$ & [10]$^{a}$ & [10]$^{a}$ & [10]$^{a}$ & [10]$^{a}$ & [10]$^{a}$ & [10]$^{a}$ & [10]$^{a}$ & [10]$^{a}$ & [10]$^{a}$ & [10]$^{a}$ & [10]$^{a}$ & [10]$^{a}$  \\      
     & $N_{\rm refl}$ & $0.105_{-0.021}^{+0.013}$ & $0.10\pm{0.03}$ & $0.10\pm{0.04}$ & $0.12\pm{0.02}$ & $0.14\pm{0.02}$ & $0.12\pm{0.03}$ & $0.16\pm{0.03}$ & $0.17\pm{0.03}$ & $0.19\pm{0.04}$ & $0.21\pm{0.05}$ & $0.19\pm{0.05}$ & $0.22\pm{0.05}$ & $0.25_{-0.09}^{+0.03}$  \\ 
     & ${F_{\rm refl}}^{*}\,(\rm{10^{-9}~ergs~cm^{-2}~s^{-1}})$ & $7.5_{-1.6}^{+1.9}$ & $6.6\pm{1.6}$ & $6.3\pm{1.3}$ & $6.3\pm{1.0}$ & $6.4\pm{0.8}$ & $5.6\pm{1.0}$ & $6.2\pm{0.9}$ & $6.7_{-0.5}^{+0.9}$ & $6.4\pm{1.0}$ & $7.6\pm{1.2}$ & $5.4\pm{0.9}$ & $5.8_{-0.6}^{+1.0}$ & $6.0_{-1.3}^{+1.6}$  \\
    gabs & $lineE$ (keV) & $0.556\pm{0.010}$ & $0.552_{-0.021}^{+0.008}$ & $0.555_{-0.016}^{+0.007}$ & $0.554_{-0.027}^{+0.008}$ & $0.553_{-0.015}^{+0.008}$ & $0.551_{-0.021}^{+0.007}$ & $0.549_{-0.023}^{+0.010}$ & $0.550_{-0.026}^{+0.008}$ & $0.548_{-0.026}^{+0.009}$ & $0.543_{-0.031}^{+0.011}$ & $0.541_{-0.033}^{+0.007}$ & $0.52\pm{0.02}$ & $0.531_{-0.028}^{+0.011}$  \\ 
     & $\sigma\,{\rm (10^{-2}\,keV)}$ & $7.0_{-0.8}^{+1.2}$ & $7.8_{-0.9}^{+2.1}$ & $7.4_{-0.8}^{+1.7}$ & $7.5_{-1.0}^{+2.2}$ & $7.7_{-0.9}^{+1.6}$ & $8.0_{-0.7}^{+1.9}$ & $8.4_{-1.0}^{+2.0}$ & $8.1_{-0.8}^{+2.2}$ & $8.7_{-0.9}^{+2.2}$ & $9.2_{-1.0}^{+2.7}$ & $9.9_{-0.6}^{+2.9}$ & $12.1_{-1.5}^{+1.6}$ & $11.0_{-0.8}^{+2.7}$  \\ 
     & Strength $(10^{-2})$  & $4.2_{-0.4}^{+0.8}$ & $4.9_{-0.5}^{+1.4}$ & $4.7_{-0.4}^{+1.0}$ & $4.7_{-0.5}^{+1.4}$ & $4.8_{-0.5}^{+0.9}$ & $5.2_{-0.3}^{+1.2}$ & $5.3_{-0.5}^{+1.4}$ & $5.1_{-0.4}^{+1.5}$ & $5.5_{-0.5}^{+1.5}$ & $6.1_{-0.6}^{+2.0}$ & $6.9_{-0.3}^{+2.4}$ & $8.8_{-1.2}^{+1.4}$ & $7.9_{-0.6}^{+2.2}$  \\ 
    vkdualdk & $kT_{\rm s,2}$ (keV) & $0.34_{-0.04}^{+0.09}$ & $0.22_{-0.03}^{+0.13}$ & $0.279_{-0.018}^{+0.112}$ & $0.34\pm{0.06}$ & $0.35_{-0.04}^{+0.07}$ & $0.39\pm{0.04}$ & $0.43_{-0.05}^{+0.08}$ & $0.48\pm{0.06}$ & $0.45\pm{0.05}$ & $0.42_{-0.05}^{+0.11}$ & $0.313_{-0.007}^{+0.132}$ & $0.305_{-0.009}^{+0.036}$ & $0.35_{-0.02}^{+0.53}$  \\ 
     & $L_{1}$ (km)  & $50_{-30}^{+80}$ & $<74$  & $27_{-10}^{+56}$ & $33_{-20}^{+33}$ & $40_{-20}^{+30}$ & $110_{-40}^{+90}$ & $70_{-30}^{+60}$ & $150_{-70}^{+180}$ & $100_{-40}^{+90}$ & $<34$ & $<38$ & $<19$ & $<95$  \\ 
     & $L_{2}\,{\rm (10^{4}\,km)}$ & $1.2\pm{0.4}$ & $1.00_{-0.54}^{+0.15}$ & $1.09_{-0.54}^{+0.12}$ & $1.1\pm{0.3}$ & $1.2\pm{0.4}$ & $2.7_{-0.7}^{+2.0}$ & $1.4\pm{0.6}$ & $2.8_{-0.7}^{+2.9}$ & $3.8_{-0.9}^{+2.3}$ & $13\pm{5}$ & $1.5\pm{0.3}$ & $1.2\pm{0.3}$ & $1.7_{-0.9}^{+3.7}$  \\ 
     & $\eta_{1}$ & $0.23_{-0.16}^{+0.09}$ & $0.51_{-0.40}^{+0.14}$ & $0.43_{-0.21}^{+0.05}$ & $0.38\pm{0.09}$ & $0.44_{-0.14}^{+0.09}$ & $0.43\pm{0.09}$ & $0.35\pm{0.14}$ & $0.33\pm{0.10}$ & $0.42\pm{0.10}$ & $0.5\pm{0.2}$ & $0.96_{-0.39}^{+0.03}$ & $1.00_{-0.08}^{b}$ & $0.82_{-0.69}^{+0.13}$  \\ 
     & $\eta_{2}$ & $1.0_{-0.4}^{b}$ & $1.0_{-0.5}^{b}$ & $0.995_{-0.565}^{+0.013}$ & $1.0_{-0.3}^{b}$ & $1.0_{-0.3}^{b}$ & $1.0_{-0.2}^{b}$ & $1.0_{-0.3}^{b}$ & $1.0_{-0.6}^{b}$ & $1.0_{-0.6}^{b}$ & $1.0_{c}^{b}$ & ${0.00_{c}^{+0.16}}$ & ${0.00_{c}^{+0.11}}$ & ${0.0_{c}^{+0.9}}$  \\ 
     & $\phi$ (rad) & $2.7\pm{0.4}$ & $2.6_{-0.4}^{+1.4}$ & $2.55_{-0.09}^{+0.61}$ & $2.44_{-0.14}^{+0.34}$ & $2.3\pm{0.3}$ & $1.6\pm{0.5}$ & $2.3\pm{0.4}$ & $1.8_{-0.7}^{+0.2}$ & $1.3\pm{0.5}$ & $-0.06_{-0.06}^{+0.19}$ & $2.48_{-0.14}^{+0.70}$ & $2.6\pm{0.3}$ & $2.6_{-1.1}^{+3.0}$ \\
     & $reflag$ & $0.14_{-0.12}^{+0.06}$ & $0.07_{c}^{+0.06}$ & $0.05_{c}^{+0.08}$ & $0.05_{c}^{+0.07}$ & $-0.11\pm{0.07}$ & $-0.15\pm{0.06}$ & $-0.18\pm{0.08}$ & $-0.26_{-0.11}^{+0.14}$ & $-0.32\pm{0.10}$ & $0.3\pm{0.3}$ & $-0.31_{-0.09}^{+0.02}$ & $-0.33\pm{0.05}$ & $-0.5\pm{0.3}$ \\
     & $\eta_{\rm{int,1}}$ & $0.05\pm{0.03}$ & $0.11\pm{0.07}$ & $0.09\pm{0.04}$ & $0.08\pm{0.02}$ & $0.10\pm{0.03}$ & $0.10\pm{0.02}$ & $0.09\pm{0.04}$ & $0.08\pm{0.03}$ & $0.11\pm{0.03}$ & $0.15\pm{0.07}$ & $0.27\pm{0.07}$ & $0.29\pm{0.01}$ & $0.2_{c}^{+0.2}$ \\
     & $\eta_{\rm{int,2}}$ & $0.21\pm{0.06}$ & $0.22\pm{0.07}$ & $0.22\pm{0.08}$ & $0.22\pm{0.04}$ & $0.23\pm{0.04}$ & $0.23\pm{0.04}$ & $0.24\pm{0.04}$ & $0.23\pm{0.09}$ & $0.25\pm{0.08}$ & $0.277\pm{0.001}$ & ${0.00_{c}^{+0.03}}$ & ${0.00_{c}^{+0.02}}$ & ${0.00_{c}^{+0.2}}$ \\
     \hline
    & $\chi^{2}/v$ & 242.7/230 & 246.0/230 & 195.2/230 & 231.0/230 & 265.6/230 & 273.7/230 & 240.3/230 & 249.1/230 & 201.9/230 & 265.1/230 & 230.5/230 & 276.8/230 & 225.9/230  \\ 
    & frequency (Hz) & $4.44\pm{0.07}$ & $4.64\pm{0.05}$ & $5.01\pm{0.05}$ & $5.50\pm{0.05}$ & $5.83\pm{0.05}$ & $6.04\pm{0.06}$ & $6.56\pm{0.11}$ & $6.34\pm{0.05}$ & $7.2\pm{0.2}$ & $7.7\pm{0.3}$ & $4.23\pm{0.04}$ & $3.98\pm{0.04}$ & $3.1\pm{0.2}$ \\
\hline
\end{tabular}
}
\begin{tablenotes} 
\footnotesize{\item[]\textit{Note.} columns and fitting limits are defined as in Table~\ref{tab:Fit_pars}. Parameter values frozen during the fitting are given in square brackets.\newline
$^{*}$ unabsorbed flux in the 0.5--10.0\,keV energy band. }
\end{tablenotes} 
\end{threeparttable}
\end{table*}

\begin{figure*}
\centering
\includegraphics[width=0.9\textwidth]{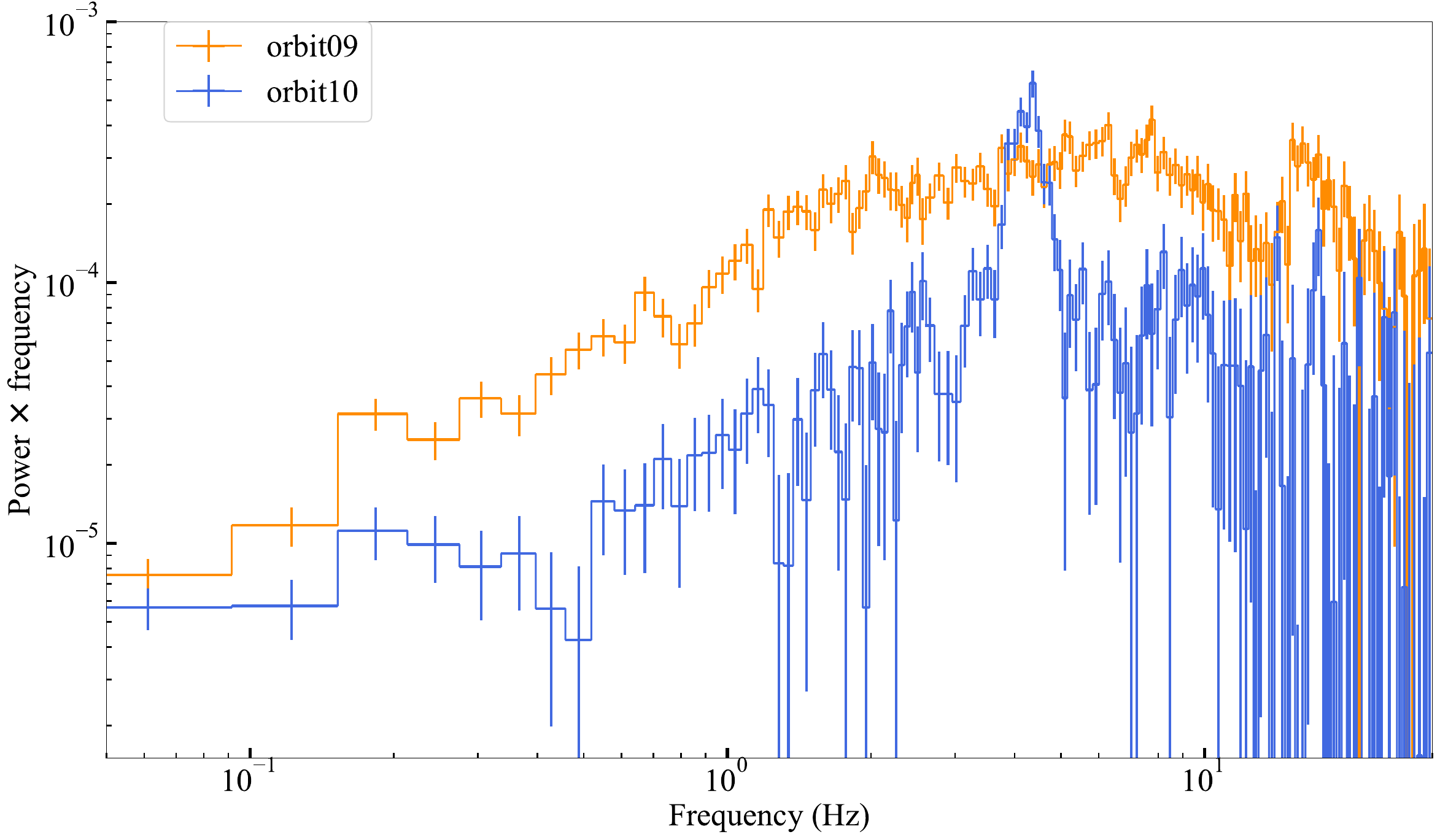} \\ 
\caption{Power density spectra in the 0.5$-$12\,keV energy band of MAXI J1820+070 with \textit{NICER}. The orange data correspond to orbit 09 in the HIMS, while the blue data correspond to orbit 10 in the SIMS. 
}
\label{fig:PDS_0910}
\end{figure*}


\bsp	
\label{lastpage}
\end{document}